\documentclass{aa}  
\usepackage{txfonts}
\usepackage{bm}
\usepackage{fontawesome}
\usepackage{hyperref} 
\usepackage{xcolor}

\newcommand{\dd}{\text{d}}
\newcommand{\Var}{\text{Var}}

\begin{document} 

   \title{Generative models of astrophysical fields \\ with scattering transforms on the sphere}

   \author{L. Mousset\inst{1}\thanks{louise.mousset@phys.ens.fr}
           \and
           E. Allys\inst{1}
           \and
           M. A. Price\inst{2}
           \and
           J. Aumont\inst{3}
           \and
           J.-M. Delouis\inst{4}
           \and
           L. Montier\inst{3}
           \and
           J. D. McEwen\inst{2}
           }
           
   \institute{Laboratoire de Physique de l’\'Ecole Normale Supérieure, ENS, Université PSL, CNRS, Sorbonne Université, Université Paris Cité, 75005 Paris, France
        \and
          Mullard Space Science Laboratory, University College London, Holmbury St Mary, Dorking, Surrey RH5 6NT, UK
        \and
              Institut de Recherches en Astrophysique et Planétologie, Université de Toulouse, CNRS, CNES, UPS, Toulouse, France
        \and
           Laboratoire d’Océanographie Physique et Spatiale, Univ. Brest, CNRS, Ifremer, IRD, Brest, France
        }

   \date{Received 08 July 2024; accepted 16 September 2024}

\abstract{Scattering transforms are a new type of summary statistics recently developed for the study of highly non-Gaussian processes, which have been shown to be very promising for astrophysical studies. In particular, they allow one to build generative models of complex non-linear fields from a limited amount of data and have been used as the basis of new statistical component separation algorithms. In the context of upcoming cosmological surveys, such as LiteBIRD for the cosmic microwave background polarisation or the \textit{Vera C. Rubin} Observatory and the Euclid space telescope for study of the large-scale structures of the Universe, extending these tools to spherical data is necessary. In this work, we developed scattering transforms on the sphere and focused on the construction of maximum-entropy generative models of several astrophysical fields. We constructed, from a single target field, generative models of homogeneous astrophysical and cosmological fields, whose samples were quantitatively compared to the target fields using common statistics (power spectrum, pixel probability density function, and Minkowski functionals). Our sampled fields agree well with the target fields, both statistically and visually. We conclude, therefore, that these generative models open up a wide range of new applications for future astrophysical and cosmological studies, particularly those for which very little simulated data is available.
}

   \keywords{Physical data and processes, Methods: data analysis, Methods: statistical, Cosmology: Large-scale structure of Universe}

   \maketitle

\section{Introduction}
\label{sec:intro}

Scattering transforms are a recently developed class of summary statistics for the study of non-Gaussian processes~\citep{mallat2012, Bruna2012}. These statistics, which are built from successive wavelet convolutions and pointwise non-linearities such as a modulus, are inspired by neural networks but do not require any training steps. Introduced recently in astrophysics~\citep{Allys2019, Allys_WPH2020}, scattering transforms have since demonstrated their ability to characterise highly non-Gaussian processes, for instance for parameter estimation and classification tasks in fields as varied as the interstellar medium~\citep{Regaldo2020Statistical, Saydjari2021classification, Lei2023}, the large-scale structures (LSSs) of the Universe~\citep{Allys_WPH2020, Cheng2020, Cheng2021, Valogiannis2022BOSS, Valogiannis2022CosmoParam}, or the epoch of reionisation~\citep{Greig2022Cosmic21cm,hothi2023wavelet}. 

Another feature of scattering transforms is that they allow one to build very efficient generative models of physical fields, in a maximum entropy framework~\citep{bruna2019multiscale}. This allows one to sample new approximate realisations of a given process relying only on its scattering transform statistics, which can be estimated from a small amount of data, sometime even a single example image~\citep{Allys_WPH2020, regaldo2023generative, price2023fast, Cheng2023}. One application of such generative models is for training data augmentation for machine learning applications. For instance, it has been shown in~\cite{Jeffrey2022}, with simulated data, that such scattering transform models constructed from a single polarised microwave dust foreground patch can be sufficient to separate primordial $B$-modes in the cosmic microwave background (CMB) polarisation from this dust emission, even in an artificially challenging mono-frequency approach. Moreover, the framework on which these generative models have been constructed has led to the development of new statistical component separation algorithms, which have for instance been successfully applied to astrophysical data~\citep{Regaldo2021denoising,JM2022,Constant2024} and seismic signals~\citep{siahkoohi2023martian, siahkoohi2023unearthing}.

While these promising scattering transform generative models have mainly been developed for 2D planar data, the adaptation of these tools to spherical data is necessary for cosmological analysis, especially for the next generation of full sky surveys, such as LiteBIRD, the Lite (Light) satellite for the studies of $B$-mode polarisation and Inflation from cosmic background Radiation Detection~\citep{PTEP_2023}, the Euclid space telescope~\citep{Euclid2011} or the \textit{Vera C. Rubin} Observatory~\citep{LSST2009}. The adaptation of a first generation of scattering transforms to spherical signals was already introduced in~\cite{mcewen:scattering} and has been used as a form of dimensionality reduction for other machine learning purposes. In this paper, we extend state-of-the-art scattering transforms~\citep{morel2023scale, Cheng2023} named the scattering covariances (SCs) to spherical fields.

The extension of scattering transforms to spherical data raises certain difficulties, namely, the definition of a directional spherical convolution with oriented filters such as wavelets~\citep{mcewen:s2let_spin, mcewen:s2let_localisation}, and the transposition of the planar translations, which appear in certain scattering transform representations. As a first step, we restricted ourselves to fields with statistically homogeneous fields with properties that do not depend on the position on the sphere. The generalisation beyond statistical homogeneity will be presented in a forthcoming paper. This naturally led us to cosmological fields, such as a weak lensing field from the LSSs of the Universe, and maps of the thermal Sunyaev-Zeldovitch (tSZ) effect of the CMB. We also consider a map of the Venus surface. In this paper, for all these spherical data, scattering transform generative models were constructed and validated from one single full-sky image.

In Sect.~\ref{sec:scatcov} we present the extension of the SC statistics to spherical fields. Then, in Sect.~\ref{sec:generative_model} we present SC-based generative models and discuss their numerical implementation. Finally, in Sect.~\ref{sec:results}, we present the results obtained with these models for the four non-Gaussian spherical fields studied. Our conclusions are presented in Sect.~\ref{sec:conclusions}.

\section{Scattering covariance on the sphere}
\label{sec:scatcov}

The SCs, or scattering spectra, were previously introduced in~\cite{morel2023scale} for 1D data and in~\cite{Cheng2023} for 2D planar maps. In this paper, we extend these statistics to spherical maps. This section introduces sampling schemes, directional convolutions, and wavelet transforms on the sphere, after which we define the SC statistics.

\subsection{Sampling of spherical maps}
\label{ss:sampling}

A spherical field can be represented by its spherical harmonic transform, which is the spherical equivalent of the Fourier transform for planar maps. In the following, we worked with the usual spherical coordinates $\bm{\omega} = (\theta, \varphi)$, with co-latitude $\theta$ and longitude $\varphi$. With these coordinates, the spherical harmonic coefficients $I_{\ell m}$ of a spherical field $I(\bm{\omega})$ defined over the sphere $\mathbb{S}^2$ correspond to the projection onto the spherical harmonics $Y_{\ell m}(\bm{\omega})$:
\begin{equation}
    I_{\ell m} = \int_{\mathbb{S}^2} I(\bm{\omega}) Y^*_{\ell m}(\bm{\omega})\dd\Omega(\bm{\omega}) \, ,
    \label{eq:Ilm}
\end{equation}
where $\dd \Omega (\bm{\omega})=\sin\theta \dd \theta \dd \varphi$ is the solid angle element.
The field can then be represented by its harmonic expansion, given by
\begin{equation}
    I(\bm{\omega}) = \sum_{\ell = 0}^{L-1} \sum_{m = -\ell}^{\ell} I_{\ell m} Y_{\ell m}(\bm{\omega}) \, .
    \label{eq:SHT}
\end{equation}
The $\ell$ index is called the multipole and is inversely proportional to angular scales on the sky, while the order $m$ at a given $\ell$ goes from $-\ell$ to $\ell$ and captures the anisotropic component of $I(\bm{\omega})$. The maximum value of $\ell$ considered, $\ell_{\rm max} = L-1$, determines the smallest scale in the transform. For a real field, the coefficients satisfy the relation
\begin{equation}
    I_{\ell m} = (-1)^m I^*_{\ell -m} \, .
    \label{eq:flm_real}
\end{equation}

The numerical computation of the forward and inverse spherical harmonic transform depends on the sampling in pixel space of the spherical map, for which different choices can be made. For example, in cosmology, the community often adopts Healpix sampling~\citep{Healpix2005}, in which all pixels have the same area, which can be an advantage in practice. With this sampling, the map resolution is given by the \texttt{nside} parameter, the number of pixels being equal to $12\times \texttt{nside}^2$. However, with this sampling, the spherical harmonic transform (as well as the Wigner transform defined below) is not accurate and needs to be refined iteratively. An alternative is instead to use an equiangular sampling, such as that defined in~\cite{mcewen:fssht}, for which these transforms can be computed exactly (to machine precision). With this sampling, abbreviated by MW, the angular dimensions of all pixels are the same, and maps are stored as 2D arrays of shape $(N_\theta, N_\varphi) = (L, 2L-1)$. In this paper, the SC statistics computations support various sampling schemes, including Healpix, MW, and others, while internal calculations typically adopt sampling schemes, such as\ MW, that afford exact spherical transforms for improved accuracy. 

Another operation required, which is also sampling dependent, is the average on the sphere, defined as:
\begin{equation}
    \langle I(\bm{\omega}) \rangle 
    =  \frac{1}{4\pi} \int_{\mathbb{S}^2}I(\bm{\omega}) \dd \Omega(\bm{\omega}).
\end{equation} 
In pixel space, this computation corresponds to
\begin{equation}
\langle I(\bm{\omega}) \rangle = \frac{1}{4\pi} \sum_p I(\bm{\omega}_p) \delta \Omega_p,
\end{equation}
where the sum is done over all pixels $p$, whose angular positions are noted $\bm{\omega}_p$. For approximate quadrature, $\delta \Omega_p$ can simply represent solid angle at pixel $p$. Alternatively, some sampling schemes exhibit exact quadrature~\citep{mcewen:fssht}, in which case $\delta \Omega_p$ denotes quadrature weights. When $I_{\ell m}$ has been computed, one directly has
\begin{equation}
    \langle I(\bm{\omega}) \rangle = \frac{1}{2\sqrt{\pi}} I_{00} \, .
\end{equation}

\subsection{Wavelet transform on the sphere}
\label{ss:wav_transform}

The SC statistics are computed from wavelet transforms, which are obtained by convolving an initial map with a set of wavelet filters, where each filter extracts the local information at a particular scale and direction. Wavelet filters need to be localised both in pixel and harmonic space. In this work, we followed~\cite{mcewen:s2let_spin, mcewen:s2let_localisation} and defined the wavelets in harmonic space in separated form as
\begin{equation}
    \Psi^j_{\ell m} = \sqrt{\frac{2\ell + 1}{8\pi^2}} \kappa_\ell^j \zeta_{\ell m}\, ,
\end{equation}
in order to control their angular and directional localisation properties separately, respectively, by kernel $\kappa_\ell^j$, with filter scale $j$, and directional component $\zeta_{\ell m}$. For the explicit definition of these two functions, one can refer to, for instance, \cite{mcewen:s2let_spin}.  The wavelets were designed to satisfy excellent spatial localisation and asymptotic non-correlation properties~\citep{mcewen:s2let_localisation}. Moreover, their directional structure in $m$ was set to zero for even/odd $m$ to enforce odd/even symmetry in $\varphi$, resulting in odd/even symmetry in $\varphi$ for $N-1$ odd/even~\citep{mcewen:s2let_localisation}.

\newcommand{\dilparam}{\eta}

\begin{figure}[t!]
    \centering
    \includegraphics[width=0.32\linewidth]{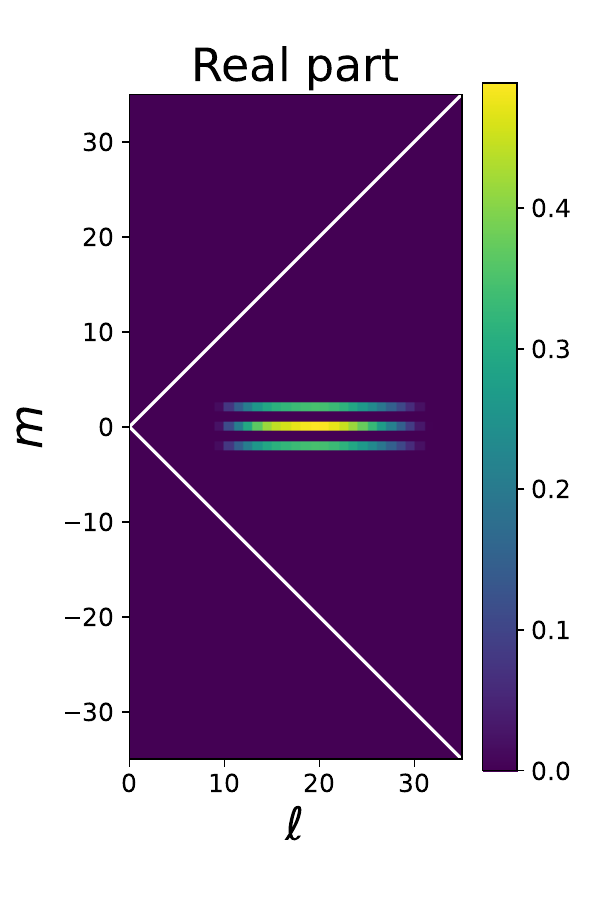}   
    \includegraphics[width=0.67\linewidth]{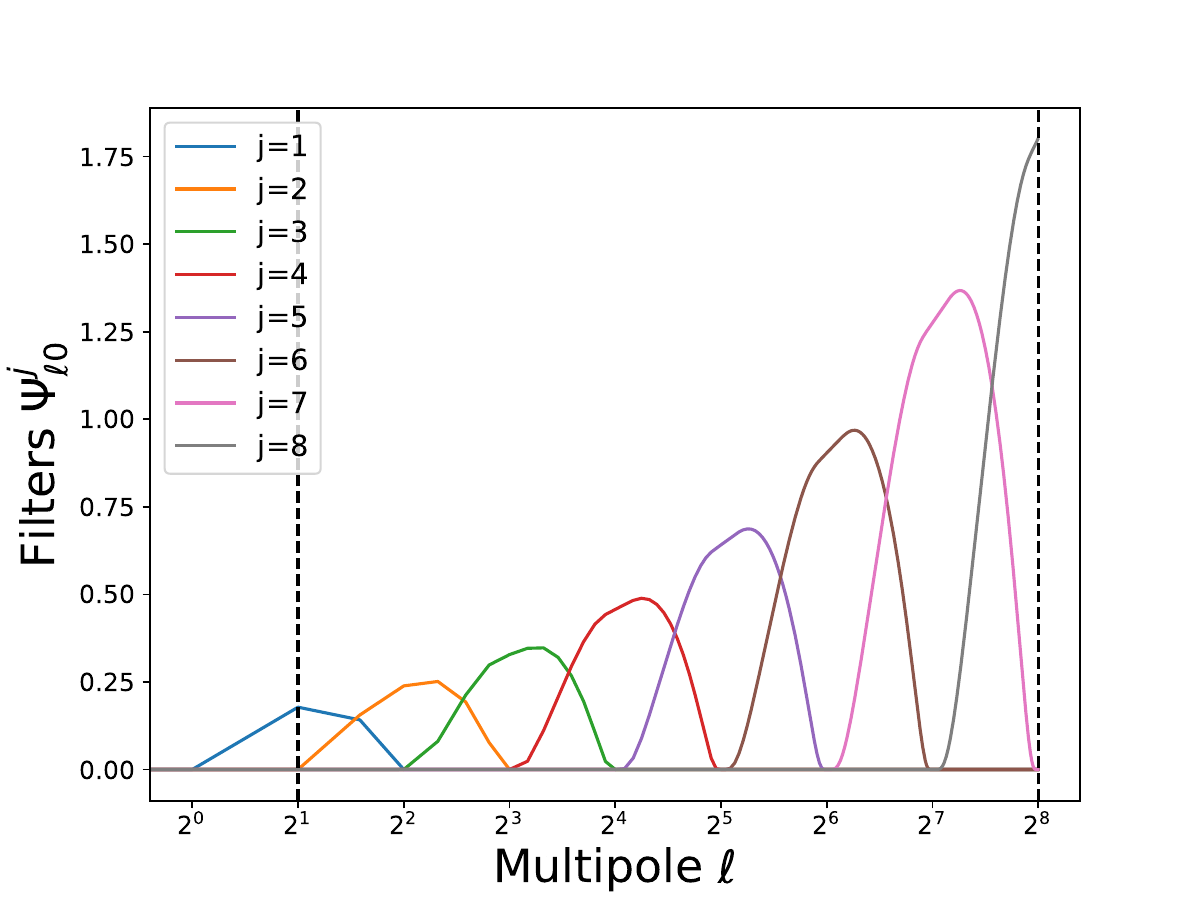}
    \caption{Harmonic representation of the wavelet filters. Left: Real part of the $\Psi^{j=4}_{\ell m}$ filter. Right: Cut at $m=0$ of the full filter set for $j$ spanning from $J_{\rm min}=1$ to $J_{\rm max}=8$.}
    \label{fig:filters}
\end{figure}

In Figure~\ref{fig:filters}, the left panel shows the $\Psi_{\ell m}$ coefficients of one filter at a specific scale $j$, and the right panel, a cut at $m=0$ of the full filter set. We note that with our convention, the $j$ scales are ordered with $\ell$ multipoles, meaning that when $j$ increases, the corresponding angular scale decreases (i.e.\ $\ell$ increases). Filters are maximum at $\ell\simeq \dilparam^j$, with support within $\ell \in [\dilparam^{(j-1)}, \dilparam^{(j+1)}]$, where $\dilparam$ defines the wavelet dilation parameter. In this paper we use dyadic scaling, corresponding to $\dilparam=2$. In the following, when performing a convolution with a filter set, the range of scales probed by the wavelets is given by $J_{\rm min} \leq j \leq J_{\rm max}$ where $J_{\rm min}\geq 0$ and $J_{\rm max} = \text{ceil}\left(\frac{\log{(L-1)}}{\log{\dilparam}}\right) $. The number of scales is given by $J = J_{\rm max} - J_{\rm min} + 1$.
The angular resolution of the wavelets is parameterised by an integer $N$, allowing for the sample of $2N-1$ directions (see below)\footnote{Although steerability could be exploited in future for further computational savings~\citep{mcewen:s2let_spin}.}.

Wavelet transforms were computed from convolutions between the field under study and this set of wavelets.  Various convolutions on the sphere can be considered~\citep{roddy:sifting_convolution}. In this work, we followed the standard directional convolution formalism presented in, for instance, ~\cite{mcewen:s2let_spin}. These convolutions produce a set of spherical maps filtered at different scales (labelled by $j$) and orientations (labelled by $\gamma$).

The directional convolution $I \star \Psi^j$ of a field $I$ with a wavelet $\Psi^j$ consists in applying a rotation by a set $\bm{\rho} = (\alpha,\beta,\gamma)$ of Euler angles of the wavelet $\Psi^j$ initially located at the north pole, before computing an inner product between the wavelet and the field $I$:
\begin{equation}
    (I \star \Psi^j)(\bm{\rho})
    = \int_\Omega  I(\bm\omega) [R_{\bm{\rho}} \Psi^j (\bm\omega)]^*  \dd \Omega,
\end{equation}
where $R_{\bm{\rho}}$ is the rotation by Euler angles $\bm{\rho}$, and $*$ stands for complex conjugation. From $(I \star \Psi^j)(\bm{\rho})$, we can identify ($\beta, \alpha$) with the spherical coordinates $\bm\omega = (\theta, \varphi)$ and $\gamma$ to the orientation that is probed in the convolution. In this way, we obtain oriented wavelet coefficients, with the shorthand notation
\begin{equation}
    (I \star \Psi^{j,\gamma})(\bm\omega) \equiv (I \star \Psi^j)(\alpha = \varphi, \beta=\theta, \gamma).
\end{equation}
While $(I \star \Psi^{j,\gamma})$ is a convenient notation, which also matches with previous work, we emphasise that there is no $\Psi^{j,\gamma}$ oriented wavelet by itself.

In practice, the directional convolution can be computed accurately and efficiently in Wigner space, which is the Fourier space associated with 3D rotations described by Euler angles. In this space, the directional convolution between a field $I$ and a wavelet $\Psi^j$ yields~\citep{mcewen:2013:waveletsxv,mcewen:s2let_spin}
\begin{equation}
    (I \star \Psi^j)^{\ell}_{m n} = \frac{8\pi^2}{2\ell+1} I_{\ell m} \Psi^{j*}_{\ell n},
\end{equation}
where $I_{\ell m}$ and $\Psi^{j}_{\ell n}$ are the spherical harmonic coefficients of $I$ and $\Psi^j$, respectively, and $(\Psi^j \star I)^{\ell}_{m n}$ are the Wigner coefficients of the convolved field, that is, the Fourier representation of the directional wavelet coefficients defined over Euler angles $\bm{\rho}=(\alpha,\beta,\gamma)$. To return to the spatial domain, we computed an inverse Wigner transform, defined as
\begin{equation}
    \label{eq:WT}
    (I \star \Psi^{j,\gamma})(\bm\omega) \equiv (I \star \Psi^j)(\bm{\rho}) = \sum_{\ell=0}^{L} \frac{2\ell +1}{8\pi^2} \sum_{m, n=-\ell}^{\ell} (I \star \Psi^j)^\ell_{m n} D_{m n}^{\ell *}(\bm{\rho}) ,
\end{equation}
where $D_{m n}^\ell(\bm{\rho})$ are the Wigner-D matrices~\citep{Varshalovich1988}. Fast (inverse) Wigner transform algorithms can then be leveraged for efficient computation~\citep{mcewen:so3,S2FFT_2023}.  By computing the wavelet transform through harmonic space as described, pixelisation artefacts are avoided. Although the wavelets satisfy an admissibility condition such that the field can be recovered exactly from its wavelet coefficients~\citep{mcewen:s2let_spin}, we are only concerned with the forward wavelet transform in this work.
In the following, we also grouped the scale and orientation under a single index $\lambda = (j, \gamma)$, writing $I \star \Psi^{\lambda}$ for the wavelet transform of the field $I$ at a given oriented scale $\lambda$.

While computing convolutions through harmonic representations is highly accurate, it involves moderate computational cost since generalised Fourier transforms on the sphere and space of rotations must be computed (albeit using fast algorithms).  An alternative would be to compute the convolutions in pixel space as done in~\cite{JM2022}. However this can introduce pixelisation artefacts.  A future avenue to consider is hybrid discrete-continuous approaches, as they have been shown to be highly computationally efficient while also avoiding discretisation artefacts~\citep{ocampo2023scalable}.

In order to optimise our numerical implementation, in particular the memory usage, all convolutions were computed in a multi-scale framework, where the map resolution is tuned to the scale at which the convolution is made. See~\cite{leistedt:s2let_axisym} for more details.

\begin{figure}[ht!]
    \centering
    \includegraphics[width=1.\linewidth]{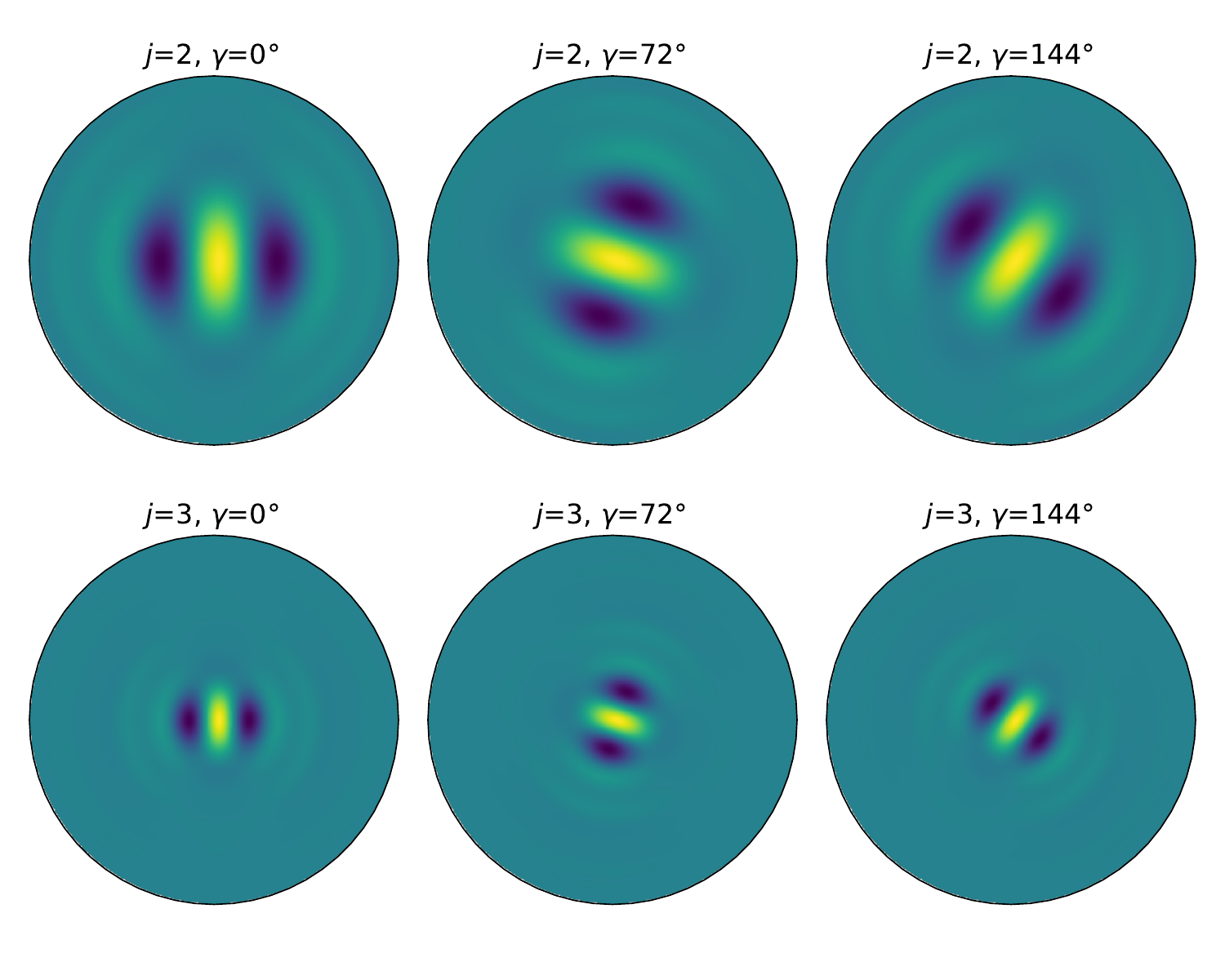}
    \caption{Spatial representation of the wavelet filters. The circles contain orthographic projections of the directional spherical wavelets for two scales and three angles, viewed looking down from the North pole. In this case, we set $N=3$ so that the angle $\gamma$ takes $2N-1 = 5$ values between 0 and $360\deg$, but only three are shown here.}
    \label{fig:conv_dirac}
\end{figure}

Figure~\ref{fig:conv_dirac} illustrates orthographic projections of the directional spherical wavelets for two scales and three angles, viewed looking down from the North pole.

\subsection{Scattering covariance coefficients}
\label{ss:scatcov_coeff}

Scattering transforms cover several types of summary statistics (see, e.g.~\cite{Allys2019, Allys_WPH2020, Cheng2023}). In this work, we consider the SCs, or scattering spectra, previously introduced in~\cite{morel2023scale} and~\cite{Cheng2023}. We chose the SCs because they only rely on convolutions and not on translations as the wavelet phase harmonic statistics~\citep{Allys_WPH2020}, which are difficult to univocally define on spherical maps.

The SC statistics characterise the power and sparsity at given scales, as well as interaction between different scales. They are built from successive applications of wavelet transforms and modulus operators, followed by average and covariance computations (assuming that we work with homogeneous processes). For a detailed introduction to these statistics, the reader is invited to refer to~\cite{Cheng2023}.
We considered two coefficients at a single scale $j_1$ and a single angle $\gamma_1$, that is, $\lambda_1 = (j_1, \gamma_1)$:
\begin{align}
    S_1^{\lambda_1} &= \langle |I \star \Psi^{\lambda_1}| \rangle \, ,\\
    S_2^{\lambda_1} &= \langle |I \star \Psi^{\lambda_1}|^2 \rangle \, ,
    \label{eq:P00}
\end{align}
and two coefficients that characterise the couplings between two and three oriented scales\footnote{
Note than in~\cite{Cheng2023}, the $S_3^{\lambda_1, \lambda_2}$ are defined as 
\[
S_3^{\lambda_1, \lambda_2} = \text{Cov} \left[I \star \Psi^{\lambda_1}, |I \star \Psi^{\lambda_2}| \right].
\]
However, as only the $\ell$ harmonics appearing in both sides of the covariance have a non-vanishing contribution, only harmonics captured by $\lambda_1$ of the $|W^{\lambda_2} I|$ term play a non-negligible role, and both formulations are closely related.
}:
\begin{equation}
    S_3^{\lambda_1, \lambda_2} = \text{Cov} \left[I \star \Psi^{\lambda_1}, |I \star \Psi^{\lambda_2}| \star \Psi^{\lambda_1}\right],
\end{equation}
\vspace{-0.7cm}
\begin{equation}
    S_4^{\lambda_1, \lambda_2, \lambda_3} = \text{Cov} \left[|I \star \Psi^{\lambda_3}| \star \Psi^{\lambda_1}, |I \star \Psi^{\lambda_2}| \star \Psi^{\lambda_1}\right],
\end{equation}
where $\langle \cdot \rangle$ corresponds to the mean over the sphere, defined in Sect.~\ref{ss:sampling}, and where the covariances are defined as $\text{Cov}[XY] = \langle X Y^*\rangle - \langle X \rangle \langle Y^* \rangle$ for two complex fields $X$ and $Y$. Note that in our case, the wavelet transforms are always zero mean, since the wavelets are mean-free. Since taking the modulus of a field mainly displaces its frequency support toward lower frequency~\citep{zhang2021, mcewen:scattering}, it is sufficient to consider terms for which $j_1 \leq j_2 \leq j_3$. Moreover, we introduce the additional parameter $\delta_j$, which corresponds to the maximum distance between pairs of scales whose interactions are characterised: that is, the calculation of $S_3$ and $S_4$ is restricted to pairs of scales such that $j_2 - j_1 \leq \delta_j$ and $j_3 - j_1 \leq \delta_j$. 

The dimension of $S_1$ and $S_2$ coefficients is $J\Theta$, with $J$ the number of scales and $\Theta = 2N-1$ the number of orientations. Regarding $S_3$ and $S_4$, their dimensions are approximately $J^2 \Theta^2$ and $J^3\Theta^3$ or $J\delta_j \Theta^2$ and $J \delta_j\Theta^3$ when considering a maximum distance between scales $\delta_j$. The exact number of coefficients are given in Table~\ref{tab:param}. 

The power spectrum of the field is mainly characterised by the $S_2^{\lambda_1}$ coefficients defined in Eq.~\ref{eq:P00}. These terms correspond to the average of the power spectrum over the $\ell$-wavelet band-passes~\citep{Cheng2023}. However, they only constrain the power spectrum over a small number of bands and this is usually not precise enough for modelling purpose. Increasing the number of scales $j$ that we probe can be done by decreasing the wavelet dilation parameter $\eta$. However, this leads to a large increase of the total number of SC coefficients. For this reason, we considered additional $S_2^{\lambda_1}$ coefficients, built with a second filter set with $\eta' < \eta$ ($\eta =2$ and $\eta'\simeq1.58$) and a single orientation $N'=1$ (isotropic filters). These coefficients are called $S_2^{\lambda_1'}$ and they allow us to constrain the power spectrum over thinner $\ell$-bins.

For physics fields the power spectra can typically be modelled by a power law, at least over certain scales~\citep{Cheng2023}. This leads to SC coefficients, which can vary over several orders of magnitude, since their amplitude is controlled by the $I \star \Psi^{\lambda} $ terms, which filters the initial $I$ field over the $j$ frequency band of the wavelet. This amplitude discrepancy can lead to ill-conditioned optimisations. Following previous works~\citep[see, for instance][]{Cheng2023}, we avoided this issue by normalising the SC statistics from the $S_2$ coefficients of a reference field that we note $S_{2, \rm ref}$. We thus define
\begin{align}
    &\bar{S}_1^{\lambda_1} = \frac{S_1^{\lambda_1}}{\sqrt{S_{2, \rm ref}^{\lambda_1}}}, \quad
    \bar{S}_2^{\lambda_1} = \frac{S_2^{\lambda_1}}{S_{2, \rm ref}^{\lambda_1}}, \quad
    \bar{S}_2^{\lambda_1'} = \frac{S_2^{\lambda_1'}}{S_{2, \rm ref}^{\lambda_1'}},\\
    &\bar{S}_3^{\lambda_1, \lambda_2} = \frac{S_3^{\lambda_1, \lambda_2}}{\sqrt{S_{2, \rm ref}^{\lambda_1} S_{2, \rm ref}^{\lambda_2}}}, \quad
    \bar{S}_4^{\lambda_1, \lambda_2, \lambda_3} = \frac{S_4^{\lambda_1, \lambda_2, \lambda_3}}{\sqrt{S_{2, \rm ref}^{\lambda_2} S_{2, \rm ref}^{\lambda_3}}} \, . 
\end{align}
When constructing generative models below, we will take the target field as the reference field, which will allows us to deal with coefficients whose values will be at most of order unity.

\begin{table}[t]
    \caption{Main simulation parameters.}
    \centering
    \tiny
    \begin{tabular}{|c|c|c|c|c|c|c|}
         \hline
         & $J$ & $J_{\min}$ & $N$ & $\delta_j$ & Trans. & Nb. of terms\\
         \hline
         LSS & 7 & 2 & 3 & 5 & log. & 8283 (35, 450, 7750) \\
         \hline
         tSZ & 6 & 3 & 3 & 5 & log. & 6173 (30, 350, 5750)  \\
         \hline
         Venus & 7 & 2 & 3 & 5 & lin. & 8283 (35, 450, 7750)\\
         \hline
         CMB & 8 & 1 & 3 & 5 & lin. & 10393 (40, 550, 9750) \\       
         \hline
    \end{tabular}
    \tablefoot{For each field, we give the number $J$ of scales that we probed, the value of $J_{\min}$, the value of $N$, which corresponds to the angular resolution of the wavelets, and the $\delta_j$ parameter, which corresponds to the maximum distance between pairs of scales whose interactions are characterised. We also indicate whether we performed a logarithmic (log.) or a linear (lin.) transformation on the target map, as discussed in Sect.~\ref{ss:maps}. The last column gives the total number of terms that compose our summary statistics $\Phi(x)$, with, in parentheses, the detailed count for $S_1$ (equal to $S_2$), $S_3$, and $S_4$.}
    \label{tab:param}
\end{table}

\section{Generative modelling}
\label{sec:generative_model}

In this section, we describe how to build generative models from the SC statistics of a given field. We also give some details on the numerical implementation of the generative models and the associated computational cost.

\subsection{Maximum entropy generative model}
\label{ss:mm_model}
We built generative models under SC constraints. These are maximum entropy micro-canonical models, which are approximately sampled by gradient descent. We refer the readers to~\cite{bruna2019multiscale} for more details. 

These models were constructed from statistics $\Phi$ estimated from a target field $x_t$; in this paper, the target field is a single full-sky map. The associated micro-canonical set $\Omega_\varepsilon$ of width $\varepsilon$ is
\begin{equation}
    \Omega_\varepsilon = \left\{ x : ||\Phi(x) - \Phi(x_t)||^2 < \varepsilon \right\},
\end{equation}
where $||.||$ is the Euclidean norm. The micro-canonical maximum entropy model is the model of maximal entropy defined over $\Omega_\varepsilon$, which has a uniform distribution over this set.

In this paper, we approximated this sampling with a gradient descent approach, which consisted in transporting a higher entropy white Gaussian distribution into a distribution supported in $\Omega_\varepsilon$. In practice, each new sample was obtained by first drawing a white noise realisation, and then performing a gradient descent in pixel space, or in harmonic space, using a loss function
\begin{equation} \label{eq:minimisation}
     \mathcal{L}(x) = ||\Phi(x) - \Phi(x_t)||^2 \,.
\end{equation} 
The typical width $\varepsilon$ of the micro-canonical ensemble was then fixed by the number of iterations used in the gradient descent. The numerical details of this implementation are given in Sect.~\ref{ss:NumExp}.

In our case, the summary statistics $\Phi(x)$ that we considered are the mean over pixels $\langle x \rangle$, its variance $\Var(x)$, and the normalised SC statistics, defined in Sect.~\ref{ss:scatcov_coeff}. Thus, we have
\begin{equation}
    \Phi(x) = \left\{ \langle x \rangle, \Var(x), \bar{S}_1^{\lambda_1}, \bar{S}_2^{\lambda_1}, \bar{S}_2^{\lambda_1'}, \bar{S}_3^{\lambda_1, \lambda_2}, \bar{S}_4^{\lambda_1, \lambda_2, \lambda_3} \right\}.
\end{equation}

We note that the target statistics $\Phi(x_t)$ were evaluated from a single full-sky image, and that the SC generative models were then built from this single set of constraints. In this respect, our approach differs from machine learning-based approaches, which generally require training on a large, and potentially very expensive, dataset.

\subsection{Details on the numerical implementation}
\label{ss:NumExp}

In this work, we considered the SC defined on the sphere, constructed using spherical wavelet transforms which in turn depend on efficient spherical harmonic and Wigner transforms (see Sect. \ref{ss:wav_transform}). As outlined in Sect. \ref{ss:mm_model}, new images were drawn from a SC micro-canonical model by minimising the loss defined in Eq.~\ref{eq:minimisation}. A plethora of algorithms have been developed to solve such optimisation problems; however, they typically require gradient information, which in turn requires that each component of the loss be differentiable.  Consequently, we required the spherical SC and thus the spherical wavelet, spherical harmonic, and Wigner transforms all to be differentiable. Recently, open-source \texttt{JAX} software that is differentiable and graphic processing unit (GPU)-accelerated has been developed for all of these transforms, including \texttt{s2fft}\footnote{\url{https://github.com/astro-informatics/s2fft}} for spherical harmonic and Wigner transforms \citep{S2FFT_2023} and \texttt{s2wav}\footnote{\url{https://github.com/astro-informatics/s2wav}} for spherical wavelet transforms \citep[][]{Price2024Differentiable}.  As part of the current work, we have developed a new open-source software implementing the spherical SC called \texttt{s2scat}\footnote{\url{https://github.com/astro-informatics/s2scat}}, which builds on top of \texttt{s2fft} and \texttt{s2wav}.

For a given target field, we computed a generated field by minimising the loss defined in Eq.~\ref{eq:minimisation} through a gradient descent in harmonic space with different initial conditions. These initial conditions are Gaussian white noises sampled in the spherical harmonic domain, that is, all of their $I_{\ell m}$ real and imaginary parts were drawn from the same Gaussian distribution such that the total variance of the target field was reproduced. In this way, the starting angular power spectrum, as defined in Eq.~\ref{eq:PS}, was flat.

To avoid a repeated spherical harmonic transform as the first step at each iteration in the computations, we chose to perform the gradient descent in the spherical harmonic domain rather than in pixel space. The variables we iterated on during the loss minimisation were thus the $I_{\ell m}$ coefficients. Because the maps are real and thanks to relation~\eqref{eq:flm_real}, we could iterate on the $I_{\ell m}$ with $m \geq 0$ only. The loss minimisation was done through a gradient descent with the L-BFGS algorithm described in~\cite{1995SJSC...16.1190B}, using the \texttt{JAX} auto-differentiable framework~\citep{jax2018github} and the \texttt{jaxopt} package~\citep{jaxopt_implicit_diff}. We stopped the optimisation after $\sim400$~iterations, which in our experiment was the typical time for the loss function to decrease by about four orders of magnitude and to reach a plateau at values around $0.1$ (meaning, since the loss was not normalised by the number of coefficients, that coefficients are on average constrained at sub-percent accuracy).

\subsection{Computational cost}
\label{ss:comput_cost}

\begin{table}[t]
    \caption{Computational benchmarking.}
    \centering
    \tiny
    \begin{tabular}{|c|c|c|c|}
         \hline
         \multicolumn{4}{|c|}{Pre-compute Mode} \\
         \hline
         Bandlimit & Forward & Gradient & JIT Compilation\\
         \hline
          256 & $15$ ms & $30$ ms & $20$ s\\      
          512 & $100$ ms & $200$ ms & $25$ s\\      
         \hline
         \multicolumn{4}{|c|}{Recursive Mode} \\
         \hline
         Bandlimit & Forward & Gradient & JIT Compilation\\
         \hline
          256 & $120$ ms & $300$ ms & $90$ s\\      
          1024 & $5$ s & $10$ s & $3$ m\\      
          2048 & $20$ s & $50$ s & $6$ m\\      
         \hline
    \end{tabular}
    \tablefoot{Results of the SC transform provided by \texttt{s2scat}. These results were recovered on a single NVIDIA A100 40GB GPU, although it is possible to run across multiple GPUs. In our analysis we generate spherical images through 400 iterations to be conservative. In practice, however, we find that $\sim 100$ iterations is typically sufficient, in which case an image at $L=256$ can be generated in $\sim 4$s. Furthermore, batched generation can dramatically decrease per sample compute time.  For example, 20 images at $L=256$ can be generated in $\sim12$s, corresponding to $\sim0.5$s per sample.} 
    \label{tab:benchmarks}
\end{table}

As outlined in Sect.~\ref{sec:scatcov}, computation of the SC statistics requires repeated spherical convolutions with subsequent non-linear activation functions, in this case modulus operators. Although directional spherical convolutions can be naively computed in pixel space with complexity $\mathcal{O}(L^5)$ \citep{mcewen:2006:fcswt}, they are more efficiently evaluated in harmonic space with complexity $\mathcal{O}(N L^3)$ \citep{mcewen:2006:fcswt,mcewen:2013:waveletsxv,mcewen:s2let_spin}. Furthermore, excellent accuracy can be achieved by computing convolutions in harmonic space since pixelisation artefacts are avoided.

We must repeatedly map to and from spherical harmonic space within our generative model using \texttt{s2fft}~\citep{price2023fast}. Two operating modes are provided by \texttt{s2scat}: one computes and caches the reduced Wigner d-functions, which are then used at runtime (pre-compute mode); and the other computes these functions on the fly through recursive algorithms (on-the-fly mode). Conceptually, the pre-compute mode is fast but requires $\mathcal{O}(L^3)$ memory, whereas the on-the-fly mode is slower but requires at most $\mathcal{O}(L^2)$ memory. When running on GPUs for harmonic bandlimits $L \leq 512$, we recommend that one adopt the pre-compute mode, deferring to the on-the-fly algorithms at higher resolutions.  Although with GPU memory increasing rapidly with hardware developments, it is likely that the pre-compute mode will be able to be run at higher bandlimits on the latest and upcoming GPUs.

High-level benchmarking results are presented in Table~\ref{tab:benchmarks}. In each case, we consider an azimuthal bandlimit of $N=3$, which corresponds to five directions on the sphere, and the full set of anisotropic SC. Our benchmarking was performed on a single NVIDIA A100 GPU with 40GB of on-board memory, although in practice \texttt{s2scat} can be distributed across a large number of GPUs. For completeness, we recorded the time for both a forward and gradient evaluation, in addition to the time required for just-in-time (JIT) compilation. One can also utilise the \texttt{jax.vmap} API, which allows one to batch calls to the maximum entropy model presented in Sect.~\ref{sec:generative_model}, resulting in more optimal GPU utilisation. For example, suppose we sample from our micro-canonical model with 100 iterations of a first-order optimiser (\emph{e.g.} ADAM; \citealt{kingma2014adam}). Generating a single new image at $L=256$ takes $\sim 4$s, whereas a batched call to generate 20 such images takes $\sim 12$s which is $\sim 0.5$s per new sample. Furthermore, the \texttt{jax.pmap} API allows one to batch calls across GPU devices, therefore accelerating generation linearly with the number of available GPUs.

\begin{figure*}[t]
    \centering 
    \includegraphics[width=1\linewidth]{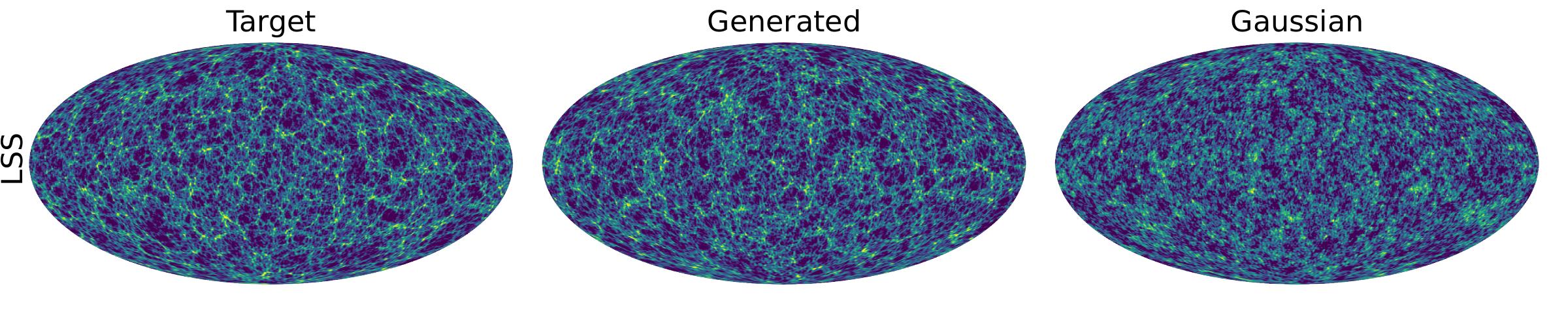}
    \vspace{-0.3cm}
    \includegraphics[width=1\linewidth]{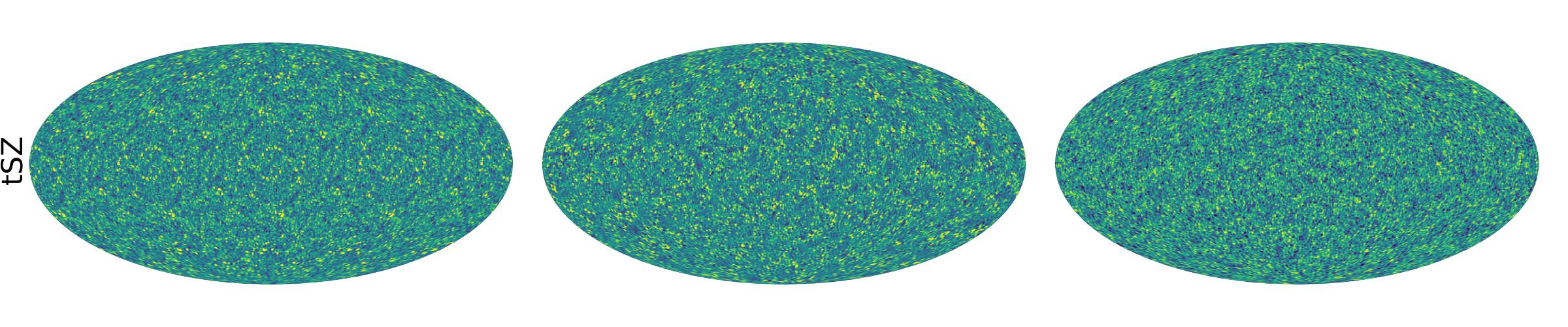}
    \vspace{-0.4cm}
    \includegraphics[width=1\linewidth]{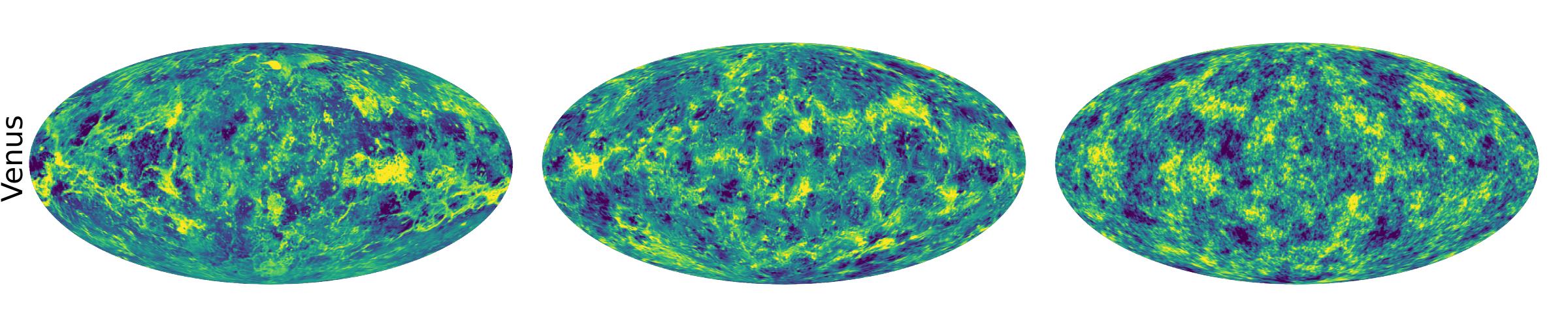} 
    \caption{Visual validation of the generative model. From top to bottom, we show the maps for the LSS, tSZ, and Venus fields. The left column is the original target field. The central column shows one sample of the generated maps. The right column shows a Gaussian field with the same power spectrum as the target. For LSS and tSZ, we plotted the logarithm of the fields in order to better see the texture details. The colour bars are identical within each field.}
    \label{fig:maps}
\end{figure*}

\section{Validation of the generative models}
\label{sec:results}

In this section, we constructed SC generative models from four astrophysical fields showing different types of structures. We then compared the generated fields with the target one. We first did a visual comparison of the maps to asses the quality of the spatial texture reproduction. We then computed various statistics in order to quantitatively evaluate our generative model. For each type of data, we drew 50~samples of the micro-canonical ensemble and computed the mean and the standard deviation over those 50~realisations. For the LSS and the tSZ fields, while the optimisation was performed on the logarithm of the maps, we compared the statistics on the raw image, after taking the inverse transform.

Before any comparison, all maps were filtered in harmonic space, keeping only the $I_{\ell m}$ coefficients such that $\ell_{\rm min} \leq \ell \leq L$ with $\ell_{\rm min} = \eta^{J_{\rm min}}$ the central frequency of the wavelet associated with $J_{\rm min}$. In this way, we considered only the scales that were constrained during the optimisation. We note however that, if necessary, it is possible to constrain in a similar scheme scales up to $\ell_\text{min} = 0$.

We also propose a comparison with samples from a Gaussian model built from the power spectrum of the target field. We produced 50~Gaussian realisations using the \texttt{synfast} method from the \texttt{Healpix} package~\citep{Healpix2005}, which can be used to construct such a Gaussian model from the angular power spectrum of a target field. This allowed us to quantify the contribution of our models built from SC statistics compared to purely Gaussian statistics. 

In Appendix~\ref{appendix_SC} we directly compared the SC statistics of the target and the generated fields. This was an additional check for the quality of the generative model. Indeed, we expected them to match as they are part of the coefficients constrained during the optimisation. 

\subsection{Description of the set of maps}
\label{ss:maps}

We first present the astrophysical and cosmological fields from which we constructed SC generative models. They were expected to have homogeneous statistical properties on the sphere. This property is essential since this is assumed when computing statistics through spatial averages. This requirement, as well as a possible way to avoid this constraint, is discussed in Sect.~\ref{ss:limit}.
The four different fields, denoted as follows, are:
\begin{itemize}
    \item{LSS}, a LSS simulation of weak lensing, from the CosmoGrid dataset~\citep{kacprzak2023cosmogridv1, PhysRevD.105.083518};
    \item{tSZ}, a tSZ effect simulated map from Simons observatory Galactic foreground simulations~\citep{Simons2019}, which were produced using the~\cite{Sehgal2010} model with modifications to better match the recent measurements;
    \item{Venus}, a map of the Venus planet from Science On a Sphere database\footnote{\url{https://sos.noaa.gov/sos/}};
    \item{CMB}, a CMB temperature map produced using the Python Sky Model software (\texttt{PySM}) from~\cite{2017MNRAS.469.2821T}.
\end{itemize}
We refer the readers to the references given above for more details on these fields. As a Gaussian field, the CMB map is a good null test for our method, as presented in Appendix~\ref{appendix_CMB}. The other fields all originate from non-linear physical processes, and thus have highly non-Gaussian structures, as can be seen in the left column of Fig.~\ref{fig:maps}. The diversity of these fields illustrates the generality and the versatility of our method, which could be used for various physical datasets.

While all of these simulated maps are available in Healpix format, the computation of the SC was done directly from the harmonic space. The conversion to this space was done using $L - 1 = \ell_{\rm max} = 2 \texttt{nside}$, and thus acts as a low-pass filter operation. This implies that spatial frequencies at $\ell \geq L$ are filtered out in this operation. We note that calculating the SC and performing the optimisation directly in spherical harmonic space means that there is no particular constraint on the sampling on the target map, even if the internal SC calculation steps are based here on MW or alternative sampling schemes to improve accuracy.

During the optimisation process, all maps were normalised such that their mean is zero and their standard deviation is one. In addition, the LSS and tSZ fields are highly non-Gaussian, making them difficult to model directly even with SC statistics. This is why we instead chose to model the logarithm of these maps\footnote{For the LSS field, the exact logarithmic transform applied on $I$ is $\log(I+\varepsilon)$ where $\varepsilon=10^{-3}$ is a regularisation to deal with non-positive fields.}. This logarithmic transform brings the distribution closer to a Gaussian one, and reduces in particular the weight of the high amplitude tail of the probability density function (PDF), allowing for better SC generative models. At the end of the optimisation, however, we took the inverse transform for these maps, and we assessed the quality of the generative model on the raw images.

The generative models were run on MW maps with a resolution of $L=256$, which corresponds to $L(2L-1) = 130816$ pixels. These real signals have $L^2 = 65 536$ complex harmonic coefficients. For the directional wavelets used to build the SC, we considered a dyadic scaling $\eta=2$ and $N=3$. This led to $J_{\rm max}=8$ and $2N-1 = 5$ orientations. As shown in Table~\ref{tab:param}, the value of $J_{\rm min}$ was tuned to each field in order to take into account the largest spatial scales that compose the maps. The number $J$ of scales that we probed is also given in the Table. The maximum distance $\delta_j$ between scales was fixed to five, which allows us to divide the total number of SC coefficients by approximately two without degrading the quality of the generative model. Concerning the additional $S_2^{\lambda'}$ coefficients, we chose an axisymmetric filter set with $N'=1$ and a scaling given by $\eta' \simeq 1.58$. The exact total number of terms making up the summary statistics $\Phi (x)$ is given in Table~\ref{tab:param}. 

\subsection{Visual validation}
\label{ss:visual_valid}

\begin{figure}[ht!]
    \centering 
    \includegraphics[width=1.\linewidth]{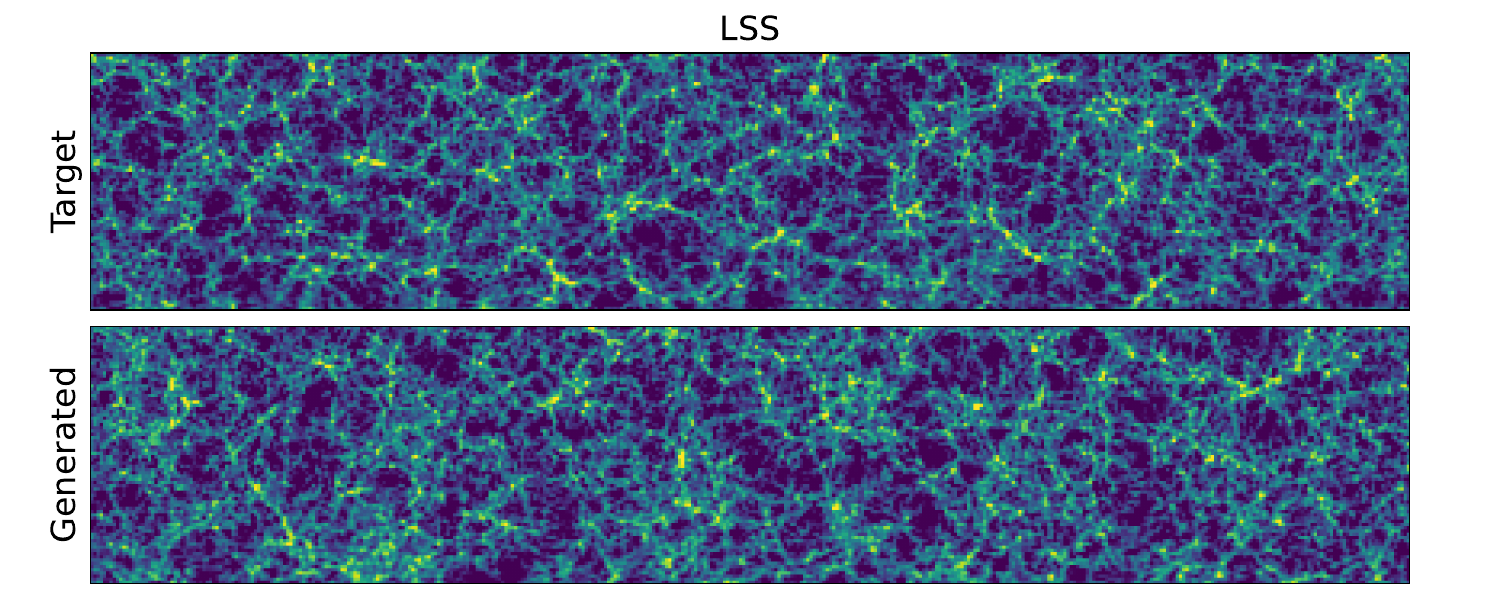}
    \includegraphics[width=1.\linewidth]{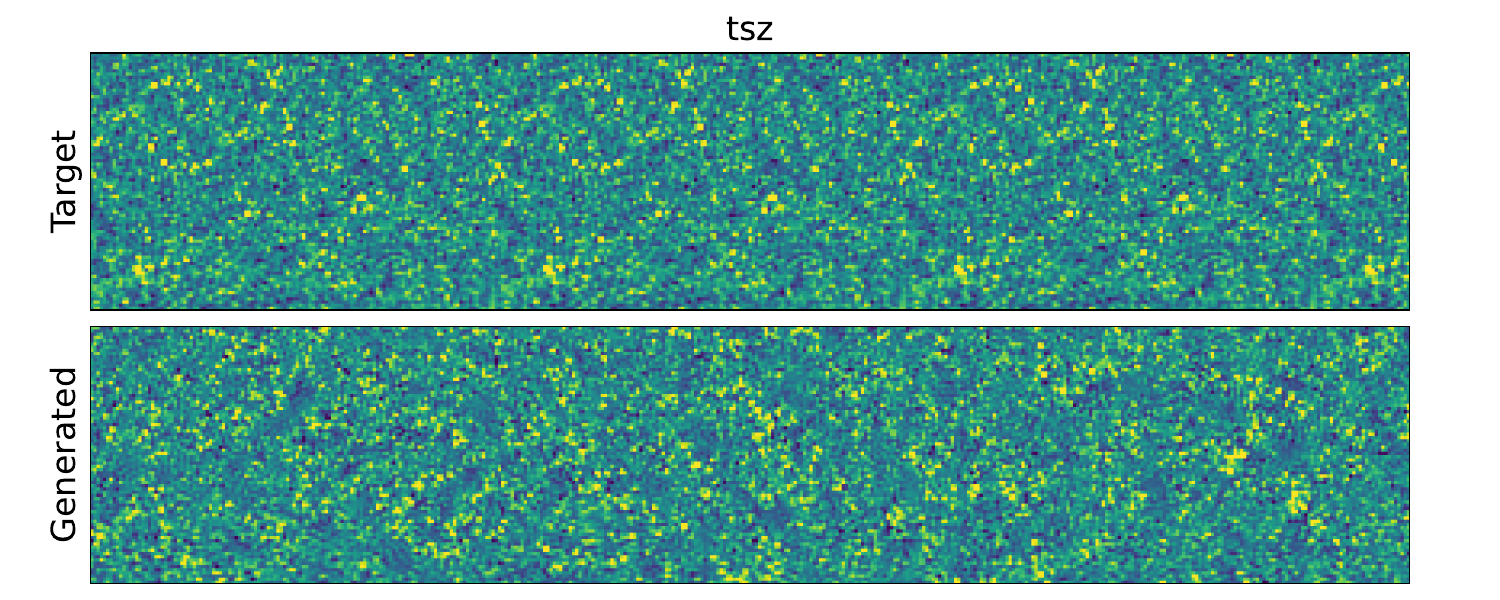}
    \includegraphics[width=1.\linewidth]{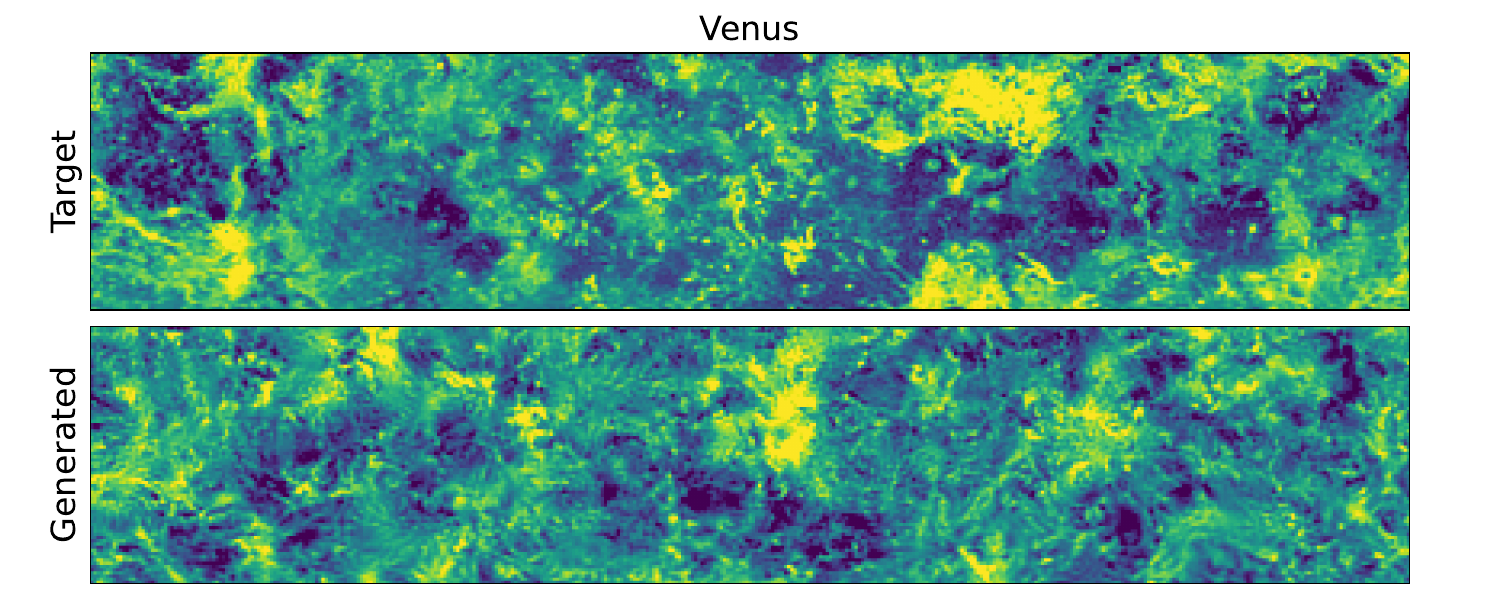} 
    \caption{Zoom-in on texture details. We show a zoom-in on a region for the LSS (top), tSZ (middle), and Venus (bottom) fields. Similarly to the previous figure, for LSS and tSZ, we plotted the logarithm of the fields in order to better see the texture details. The colour bars are identical within each field.}
    \label{fig:maps_zoom}
\end{figure}
As a first test, we can visually compare the target field and the generated ones, as presented in Fig.~\ref{fig:maps} for the LSS, tSZ, and Venus fields. They appear to be visually very similar to the original maps, which clearly shows that the SC statistics capture an important part of the non-Gaussian texture of the field. On the contrary, the structures are not reproduced in the Gaussian realisations shown on the right column. For LSS and tSZ, we plot the logarithm of the fields, which allows us to better see the textures. In addition, in Fig.~\ref{fig:maps_zoom}, we show a zoom on a smaller region to better visualise the details in the spatial structures. In Appendix~\ref{appendix_Nreal}, we also show four realisations of the fields starting from different initial conditions. This shows the ability of our generative models to sample independent realisations while capturing the overall texture of the fields.

\begin{figure*}[t]
    \centering
    \includegraphics[width=0.32\linewidth]{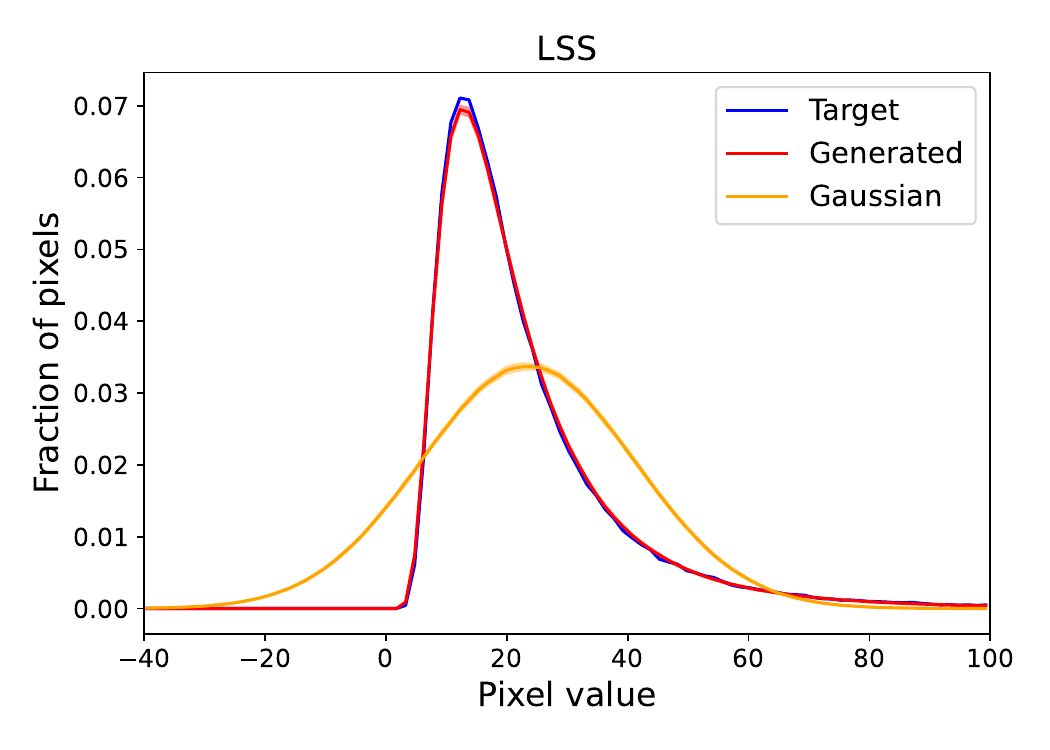}   
    \includegraphics[width=0.32\linewidth]{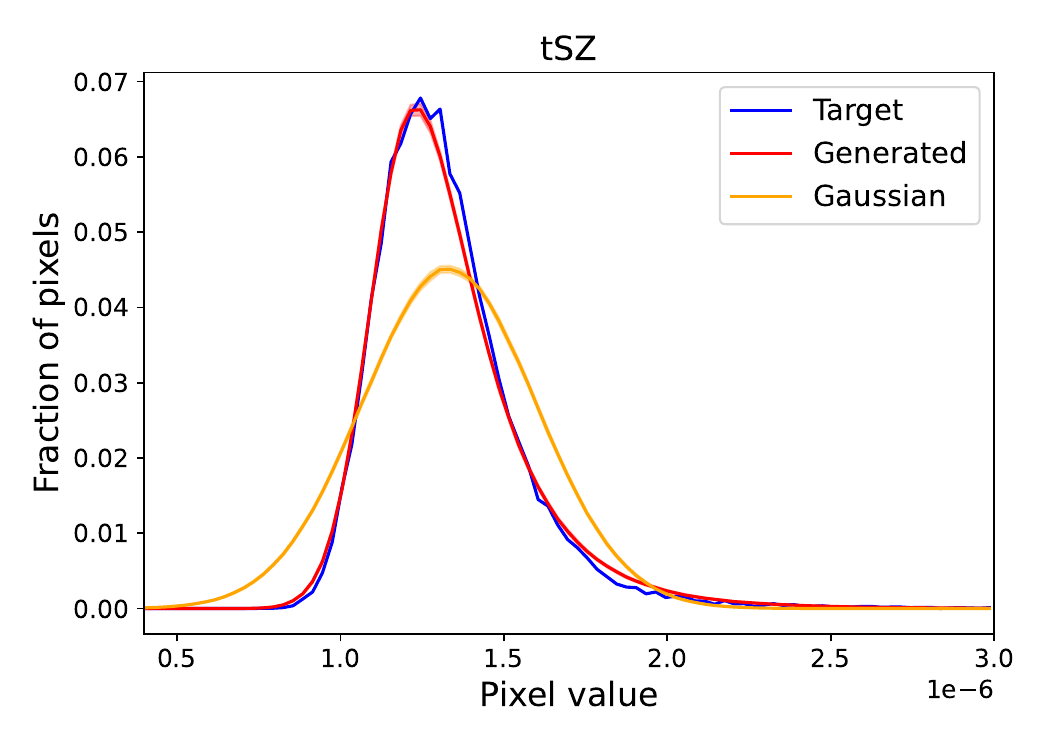}
    \includegraphics[width=0.32\linewidth]{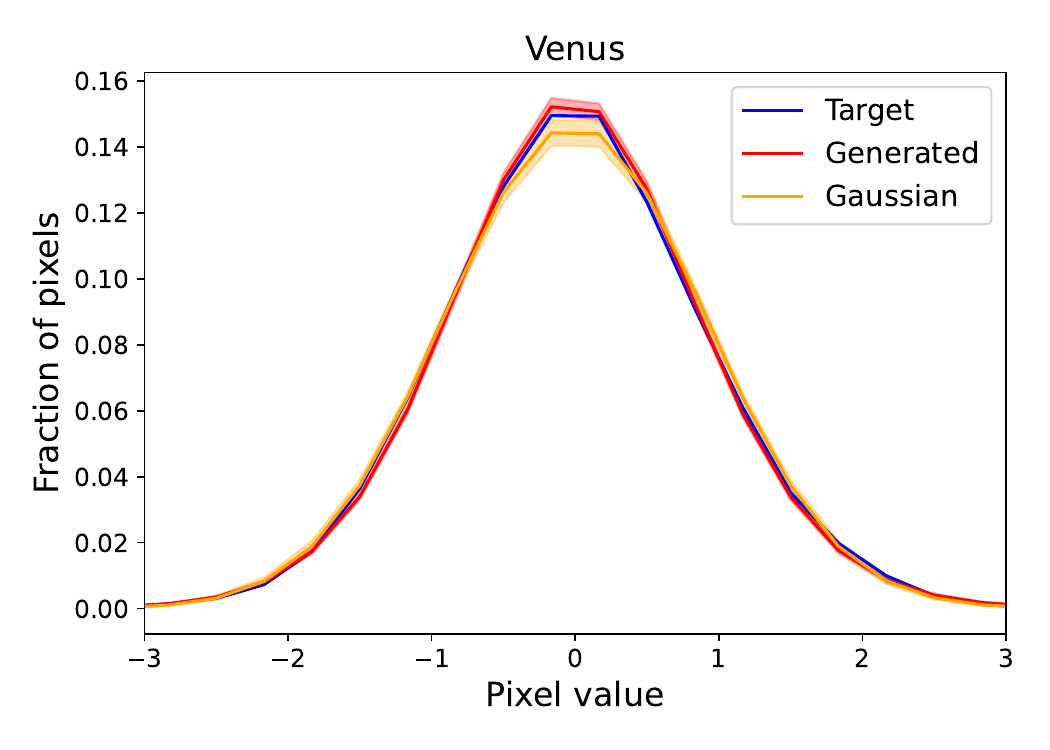} 
    \includegraphics[width=0.32\linewidth]{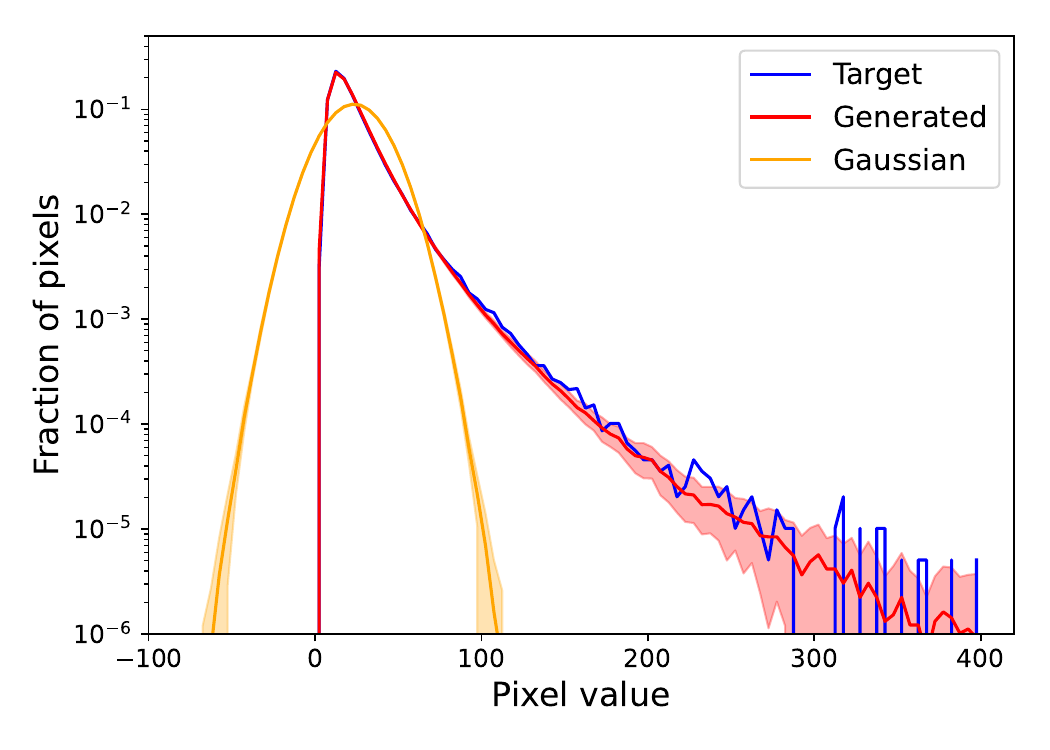}   
    \includegraphics[width=0.32\linewidth]{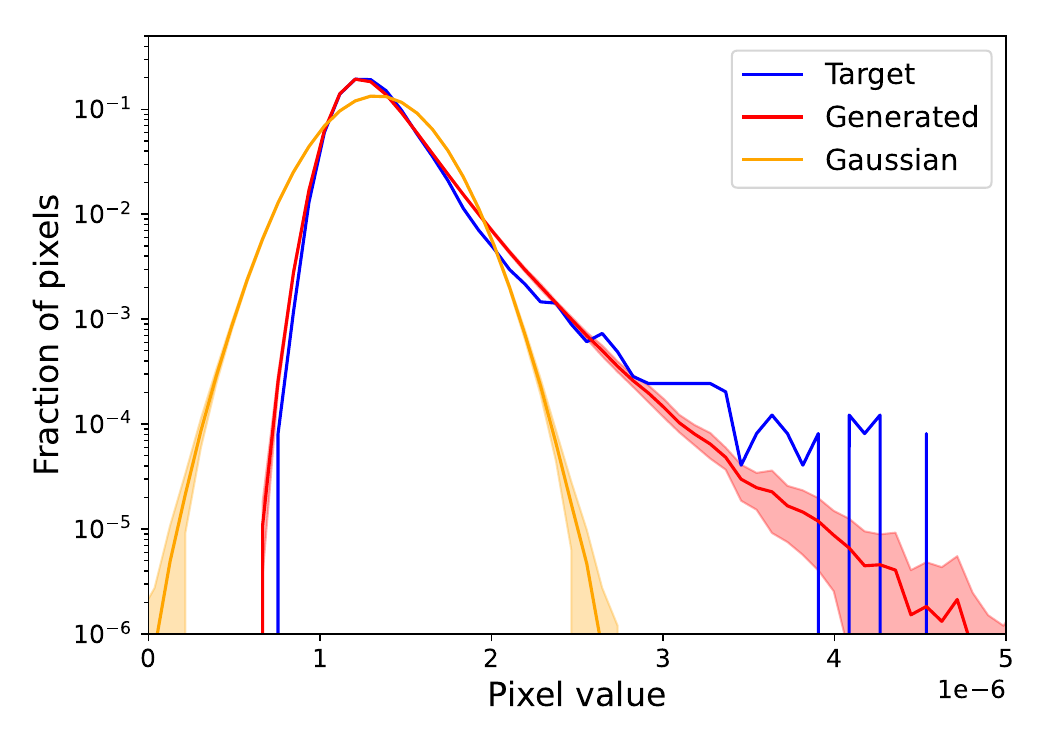}
    \includegraphics[width=0.32\linewidth]{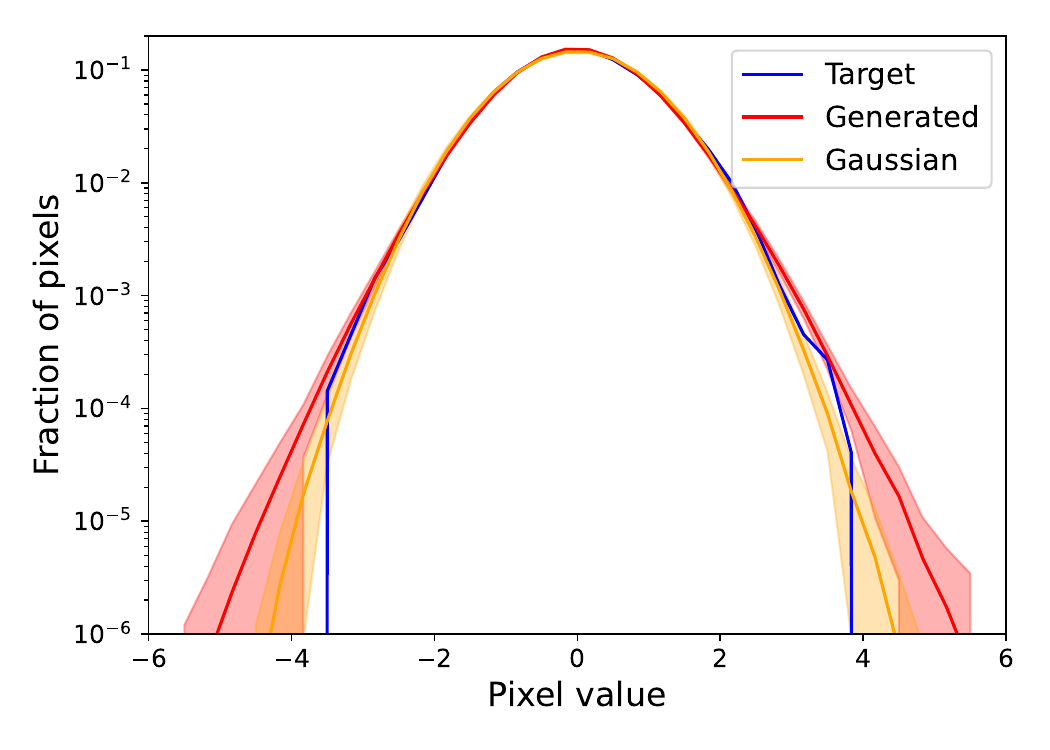}
    \includegraphics[width=0.32\linewidth]{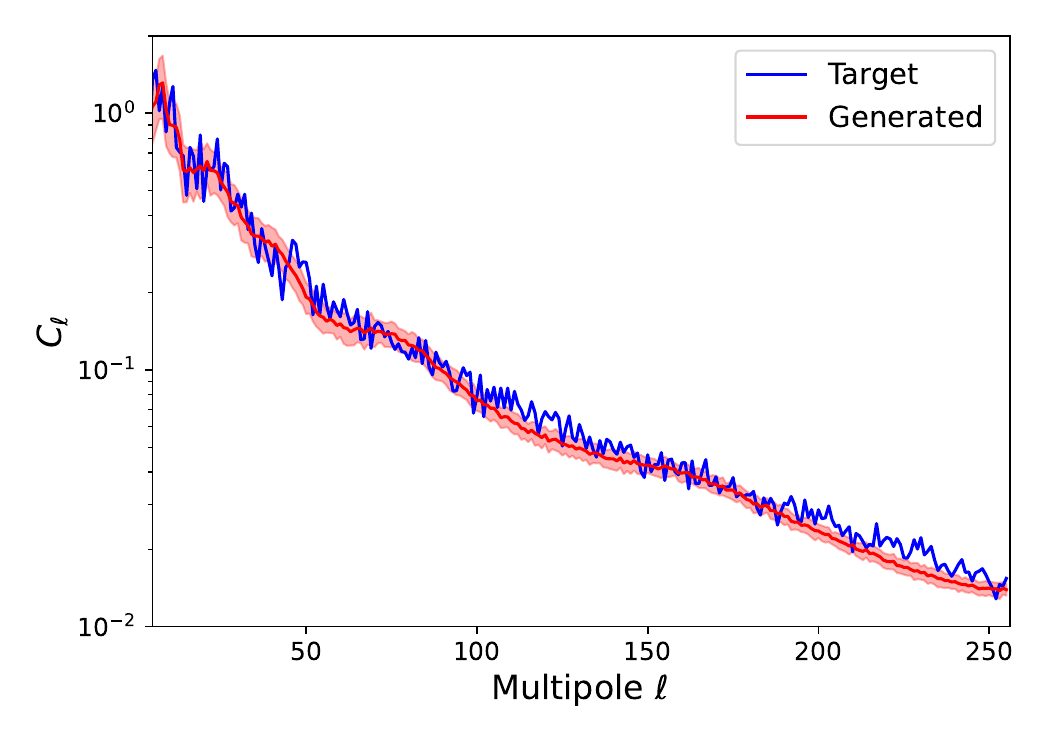} 
    \includegraphics[width=0.32\linewidth]{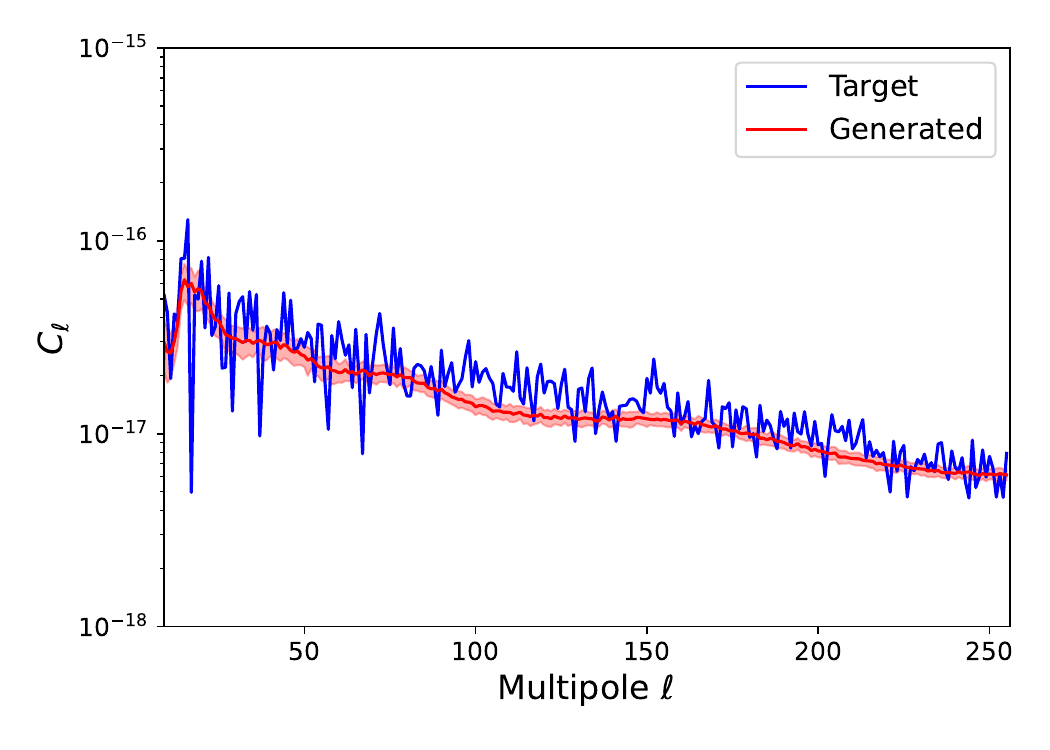}  
    \includegraphics[width=0.32\linewidth]{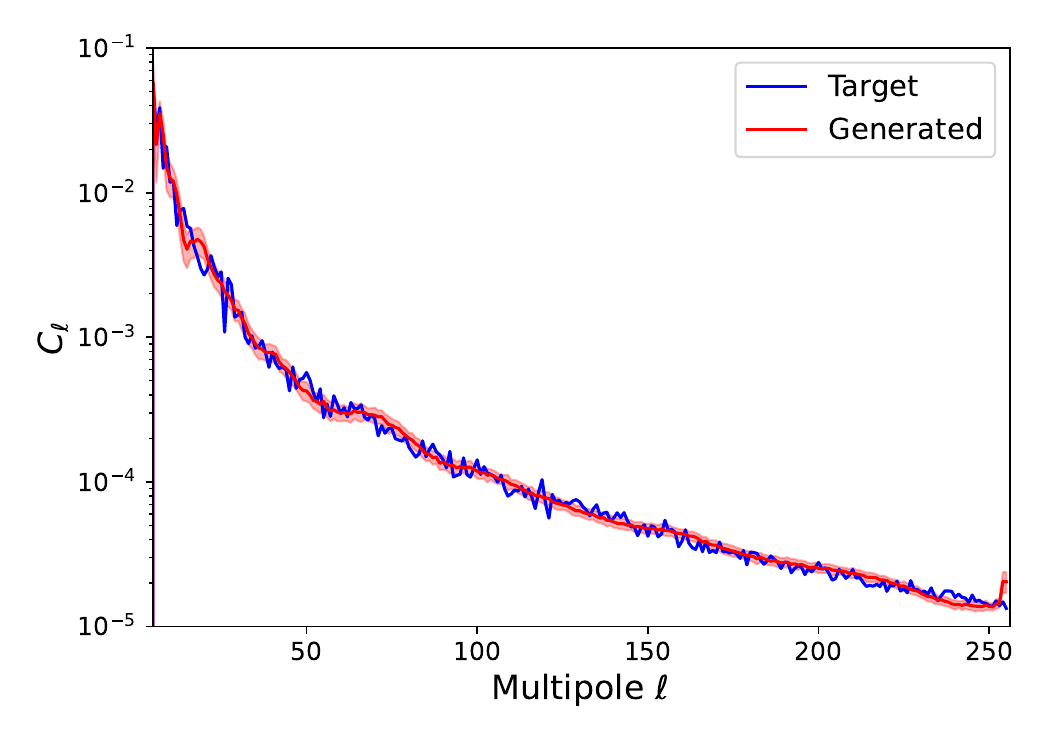}   
    \caption{Statistical validation. The PDF and angular power spectra for the LSS, tSZ, and Venus fields (left to right). The first row shows the PDF with a linear y-axis scaling, while the second row shows the same PDF with a logarithmic y-axis. The third row shows the angular power spectra. The target is shown in blue, the generated fields in red, and the Gaussian realisations in yellow. We plotted the mean (solid line) and the standard deviation (shadow envelope) over 50~realisations.}
    \label{fig:pdf_ps}
\end{figure*}

\subsection{Statistical validation with standard summary statistics}

Following a similar approach to~\cite{Cheng2023} and~\cite{price2023fast}, we compared summary statistics between the target field and the generated field. As previously, we show the mean and the standard deviation computed over 50~realisations. The summary statistics we chose to compare are:
\begin{itemize}
    \setlength\itemsep{0.2em}
    \item the PDF of the map;
    \item the angular power spectrum; 
    \item the three Minkowski functionals.
\end{itemize}

The PDF of the maps and the three Minkowski functionals were performed on Healpix maps. This was done by projecting the output $I_{\ell m}$ from the loss minimisation onto the Healpix map by an inverse spherical harmonic transform at the end of the generative process.

\subsubsection{Probability density function}

The PDF for the LSS, tSZ, and Venus fields are shown in Fig.~\ref{fig:pdf_ps}, computed on the Healpix maps. On the first row, we show the PDF with a linear y-axis scaling, while on the second row, we show them with a logarithmic y-axis scaling in order to better exhibit the tails of the distributions on several orders of magnitude. The target fields are shown in blue and the generated ones in red. In yellow, we also show the comparison with the Gaussian realisations. By definition, the PDF of these realisations presents a Gaussian profile in linear scale and a parabolic profile in logarithmic scale. 

While the Venus field has a PDF that only slightly differs from the Gaussian case, the two other fields clearly have non-Gaussian features with large tails. The comparison of the target and generated PDFs with the Gaussian PDF also allows us to better see their non-symmetric shape, which is characteristic of non-Gaussian features. As we can see, the PDFs for SC models are well reproduced on at least three orders of magnitude. The results obtained for the LSS fields, which are very good up to five orders of magnitude, are especially striking. On the other hand, the results for the tSZ field begin to push the expressive limit of our generative models. For the Venus maps, we have identified the abrupt jump in the histogram as a flaw in the data used, which does not particularly illustrate a limitation of our maximum entropy SC model.

\subsubsection{Angular power spectrum}

We calculated the power spectrum in the usual way as
\begin{equation}
    C_{\ell} = \frac{1}{2\ell + 1} \sum_{m=-\ell}^{m=\ell} |I_{\ell m}|^2 \, ,
    \label{eq:PS}
\end{equation}
where the normalisation factor $1/(2\ell + 1)$ yields a flat power spectrum in case of white Gaussian noise.

As discussed in Sect.~\ref{ss:scatcov_coeff}, the power spectrum was constrained during the optimisation through the $S_2$ and the additional $S'_2$ coefficients. We note however that these coefficients did not constrain the full power spectrum, as each term constrains only a weighted power spectrum over the frequency support of the associated wavelet. 

The third row of Fig.~\ref{fig:pdf_ps} shows the results for the generation of the LSS, tSZ, and Venus fields, from left to right. Power spectra are well reproduced over all scales, even when they vary over up to four orders of magnitude. However, small oscillations around the target can be seen in the generated power spectra. These are residual features related to the frequency bands of the wavelets, which illustrate the trade-off between the quality of reproduction we want to achieve and the number of filters we use, that is to say, the computational efficiency of our generative model.

We note that in this paper, we include 11 of these additional $S'_2$ coefficients. This number, and the precise shape of the wavelets used, could however be tuned to better reproduce the power spectrum of the target. However, care must be taken not to over-constrain these terms, as all of the samples generated would then have a power spectrum very close to that of the target, which does not necessarily correspond to a good generative model. The introduction of $S'_2$ terms is an improvement with respect to previous work, since it allows us to have better power spectrum constraints without significantly increasing the overall number of SC statistics.

\subsubsection{Minkowski functionals}

Finally, we computed the Minkowski functionals on the Healpix maps. These standard non-Gaussian statistics characterise the topology of the level sets of the field. In two dimensions, there are three Minkowski functionals, $V_0(u)$, $V_1(u)$, and $V_2(u)$, that depend on a pixel value threshold $u$. We computed them using \texttt{Pynkowski} software~\citep{Pynkowski2024}. We refer the reader to this publication for the complete definition of those statistics. The result is shown in Fig.~\ref{fig:Minkowski} only for the LSS field, while the others are shown in Appendix~\ref{appendix_Minko}. In yellow, as a comparison, we show the case of the Gaussian realisations. For the generated fields and the Gaussian realisations, we plotted the mean (solid) and the standard deviation (shadow envelope) computed over 50~realisations. Thus, the SC models encompass very well these non-Gaussian statistical features.

\begin{figure}[t!]
    \centering
    \includegraphics[width=1.\linewidth]{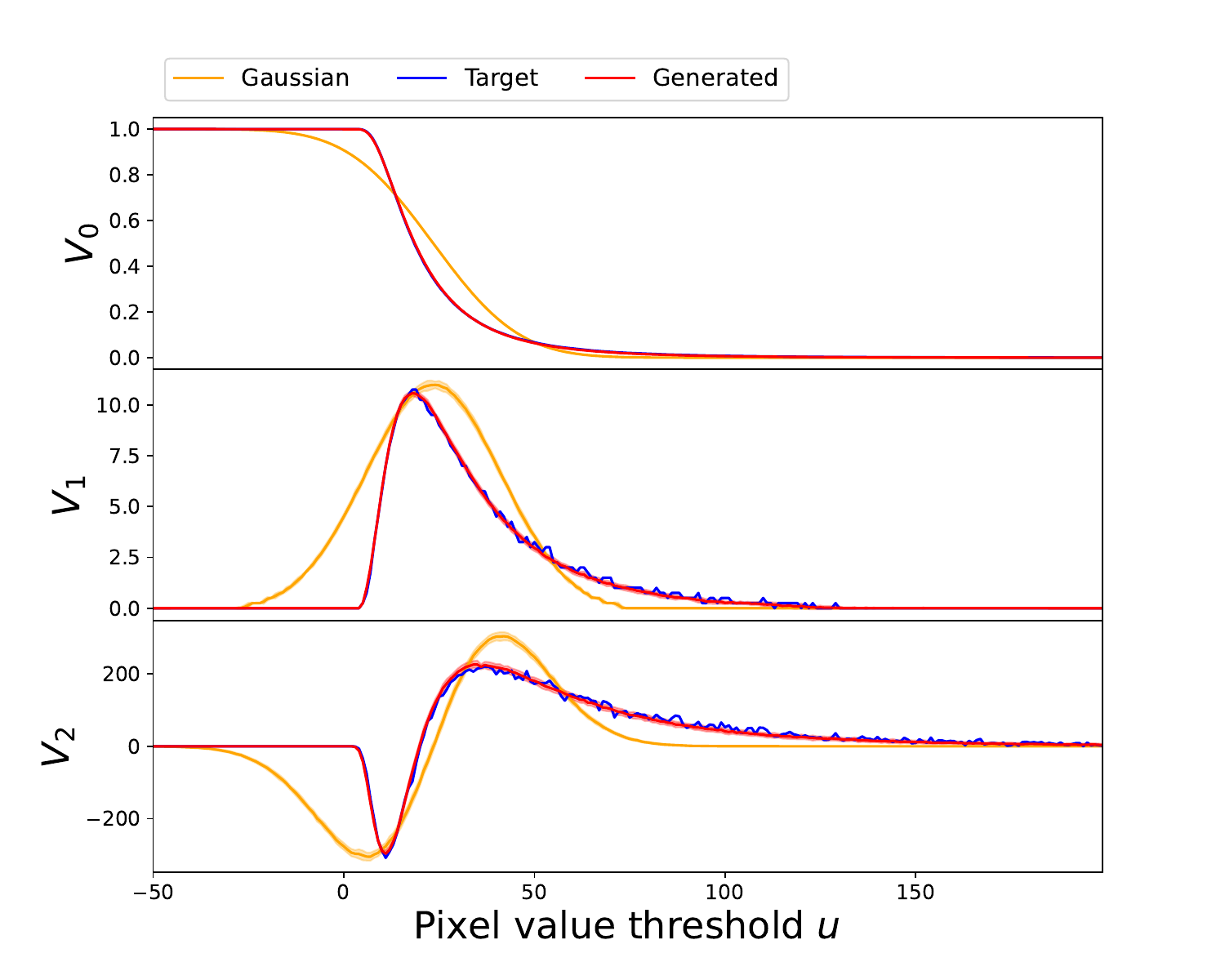}    
    \caption{Minkowski functionals. We plotted the three Minkowski functionals $V_0$, $V_1$, and $V_2$ for the LSS field. Blue is the target, red the generated fields, and yellow the Gaussian fields. For the generated fields and the Gaussian realisations, we plotted the mean (solid) and the standard deviation (shadow envelope) computed over 50~realisations.}
    \label{fig:Minkowski}
\end{figure}

\subsection{Limitations and discussions}
\label{ss:limit}

This work is a first implementation of generative models from state-of-the-art spherical SCs. As a first proof of concept, we constructed and validated these models on various cosmological fields, most of them simulated. However, building such models from real data, or using these tools to perform statistical component separation, may require some additional work to deal with their own specificity. In this section, we comment on some of these limitations and how to overcome them.

A first limitation of our current work is the fact that our models assume the statistical homogeneity of the fields studied. However, the ability to deal with non-homogeneous physical processes is usually required when modelling astrophysical fields, such as the Galactic emissions, whose properties typically vary strongly with latitude on the sky. An efficient way to deal with this issue is to rely on different masks in pixel space, for which statistical properties can be constrained independently~\citep{JM2022}. However, this requires a trade-off between the size and number of masks: using a larger number of masks gives a better description of large-scale variations in statistical properties but increases the variance of the SC statistics estimates on each mask due to the smaller number of pixels used, in addition to increasing the total number of SC coefficients and the computational and memory cost.

A second limitation is the map resolution that we can achieve. For now, the generation of a new map at $L=256$ and $N=3$ takes $\leq 1$s on a single GPU. This is thanks in part to a large number of pre-computed matrices necessary for the Wigner transform, which are stored in memory (several Gigabytes). Specifically this memory is cubic with $L$, which is prohibitive at high $L$. When increasing the resolution beyond $L \sim 1024$, we usually reach the GPU memory limit and the coefficients have to be computed on the fly, although this of course depends on available GPU specifications. Critically, the on-the-fly approach dramatically reduces memory requirements so that generations at high $L$ are at least feasible but at the cost of a significantly increased computation time. In future work, we will explore further optimisations for high $L$. A key avenue we will explore is the introduction of hybrid wavelet convolutions, which operate efficiently in pixel and harmonic space at high and low resolutions, respectively \citep[see \emph{e.g.}][]{JM2022, ocampo2023scalable}.

\section{Conclusions}
\label{sec:conclusions}

The main result of this paper is the extension of state-of-the-art scattering transforms to spherical fields. We have worked with the last generation of scattering transform statistics, named SCs~\citep{Cheng2023}, which were previously introduced for 1D and 2D planar fields. They have the advantage of relying only on successive wavelet transforms and modulus, as well as on covariances, and do not require any translations. We have also used state-of-the-art directional wavelets on the sphere, computed in spherical harmonic space~\citep{mcewen:s2let_spin}. The numerical implementation of this work, \texttt{s2scat}, is open-source and publicly available. Furthermore, it is fully auto-differentiable, using the JAX Python framework~\citep{jax2018github} and building on the \texttt{s2fft}/ \texttt{s2wav} packages~\citep{S2FFT_2023, Price2024Differentiable}.

These developments allow us to build generative models of full sky spherical fields without the need for large training datasets. In fact, our method holds in the limit case of a single data realisation. The performance of those generative models was validated quantitatively on different fields: a LSS weak lensing field, maps of the tSZ effect and of the CMB, and a map of the Venus surface, for which they performed extremely well. The diversity in terms of structures between the maps shows the impressive ability of SCs to comprehensively characterise very different non-Gaussian textures. 

This work introduces a new powerful innovative approach for spherical data, and it opens interesting perspectives for astrophysical applications. In particular, we plan to use it for the study and the modelling of CMB astrophysical foregrounds. The first goal will be to have a tool to produce multiple realisations of the different astrophysical components, for example, using the AGORA simulations~\citep{Agora2022}. Then, scattering transforms could play a role in component separation, relying on both recently developed scattering transform-based statistical component separation approaches, as well as investigating how classical component separation methods could benefit from scattering transforms, using the non-Gaussianities as an additional lever arm for disentangling different components.  

Finally, we also point out that SC statistics provide highly informative sets of statistics, which could be very useful in tasks such as parameter inference, such as simulation-based inference (SBI; \citealt{cranmer2020frontier}), from large cosmological surveys~\citep[see, for instance][]{regaldo2024galaxy,gatti2024dark,cheng2024cosmological}. This could be all the more useful as the compression factor they enable, compared with a direct description in pixel space, becomes extremely large at high resolution, due to logarithmic scale binning. This property could be further enhanced by using compression schemes such as the one presented in~\cite{Cheng2023}, which can make SCs a very informative and versatile compressed set of statistics.

\paragraph{Data availability:}
We make our code available to the community so that this work can be easily reproduced and developed further. \href{https://github.com/astro-informatics/s2scat}{\faGithub}

\begin{acknowledgements}
    We thank Sixin Zhang and Anthony Banday for very helpful discussions all along the project. Most of the simulations were produced on the MesoPSL calculation centre, for which we thank the administrators. This work was supported in part by a CNES postdoctoral fellowship. MAP and JDM are supported in part by EPSRC (grant number EP/W007673/1) and STFC (grant number ST/W001136/1).
\end{acknowledgements}

\bibliographystyle{aa} 
\bibliography{scat} 

\appendix

\section{SC statistics}
\label{appendix_SC}

\begin{figure}[ht!]
    \centering
    \includegraphics[width=1.\linewidth]{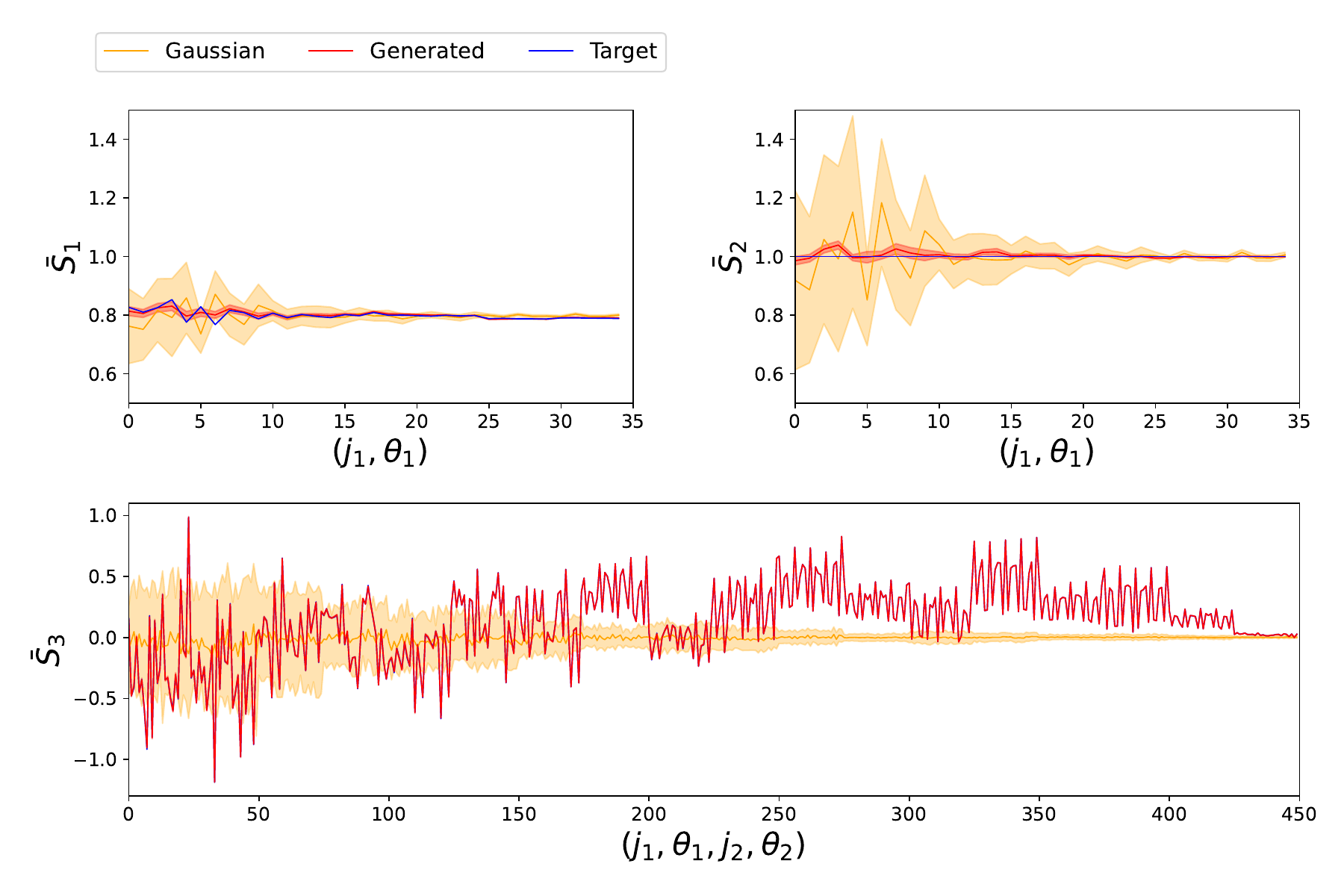}
    \caption{Normalised SC coefficients. We plot $\bar{S}_1$, $\bar{S}_2$, and $\bar{S}_3$ for the logarithm of the LSS field. We show the coefficients from the target field (blue), the generated fields (red), and the equivalent Gaussian realisations (yellow). The mean and the standard deviation over 50~realisations are shown as a solid line with a shadow envelope.}
    \label{fig:scat_coeff}
\end{figure}

In Figure~\ref{fig:scat_coeff} we show the normalised SC coefficients $\bar{S}_1$, $\bar{S}_2$, and $\bar{S}_3$ of the LSS field. We plot the coefficients of the target field in blue, the generated ones in red, as well as the Gaussian realisations in yellow. Coefficients are plot following the lexicographic order. We chose not to show $\bar{S}_4$ for readability because of the large number of coefficients.

Regarding the Gaussian realisations, shown in yellow, we expect the $\bar{S}_3$ coefficients to be equal to zero up to the correlations induced by the overlapping between wavelet bands. As we can see, for $\bar{S}_3$, the mean is centred on zero.

By construction, $\bar{S}_2$ for the target field is equal to one. This is because we have considered the ${S}_2$ coefficients of the target as the reference to normalise all the coefficients, as described in Sect.~\ref{ss:scatcov_coeff}. In this way, all the normalised coefficients are of the order of the unit. As we can see, SC statistics are well constrained by the optimisation, the generated coefficients in red well overlap the target coefficients in blue. $\bar{S}_3$ coefficients strongly differ from the Gaussian field, showing clear non-Gaussian signatures. 

\section{CMB map as a null test}
\label{appendix_CMB}

\begin{figure*}
    \centering
    \includegraphics[width=1\linewidth]{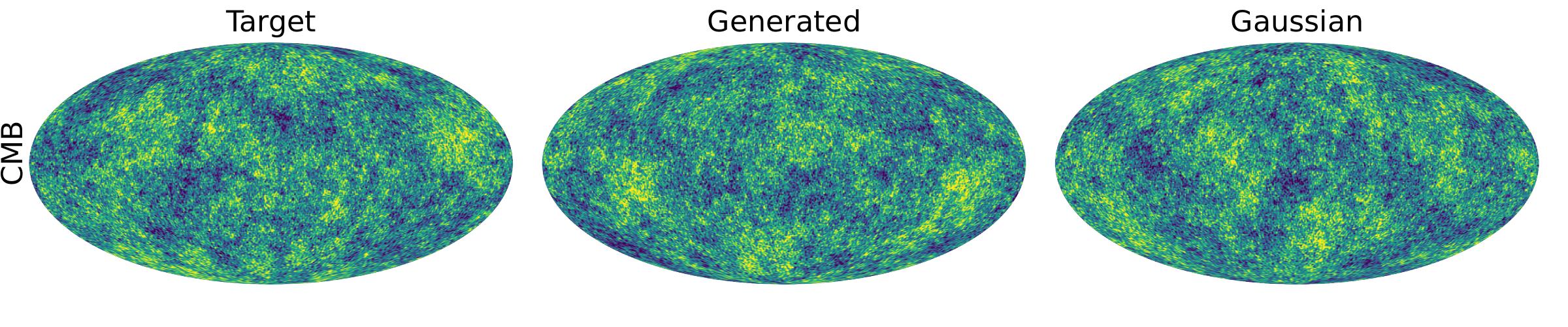}
    \includegraphics[width=0.32\linewidth]{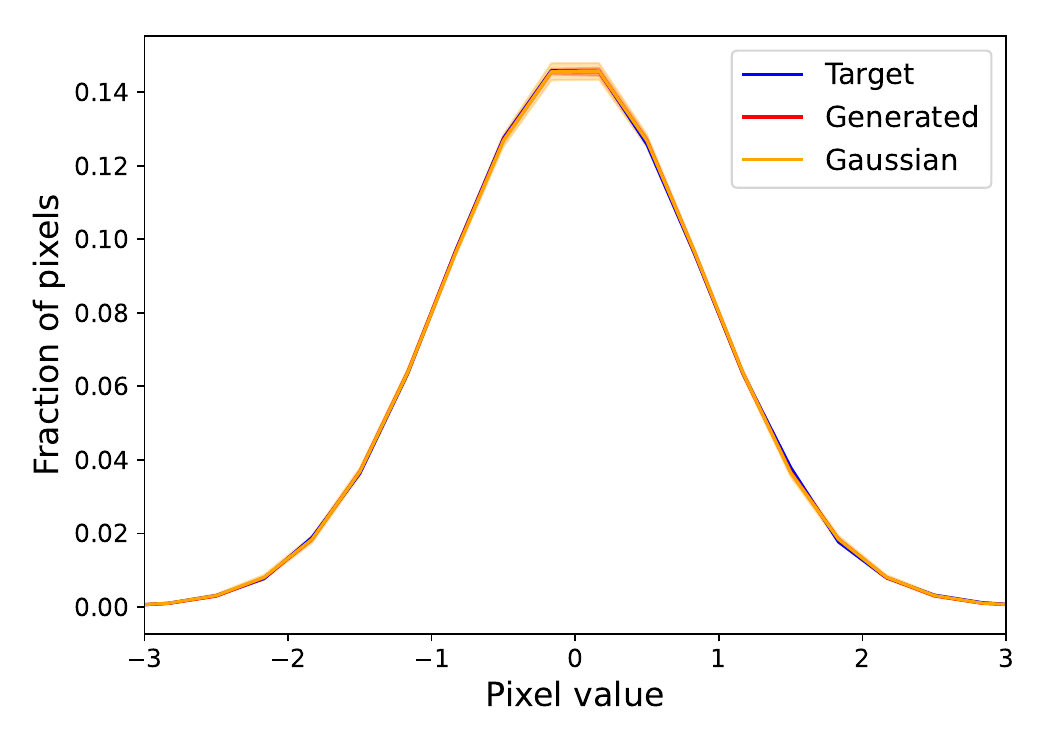}
    \includegraphics[width=0.32\linewidth]{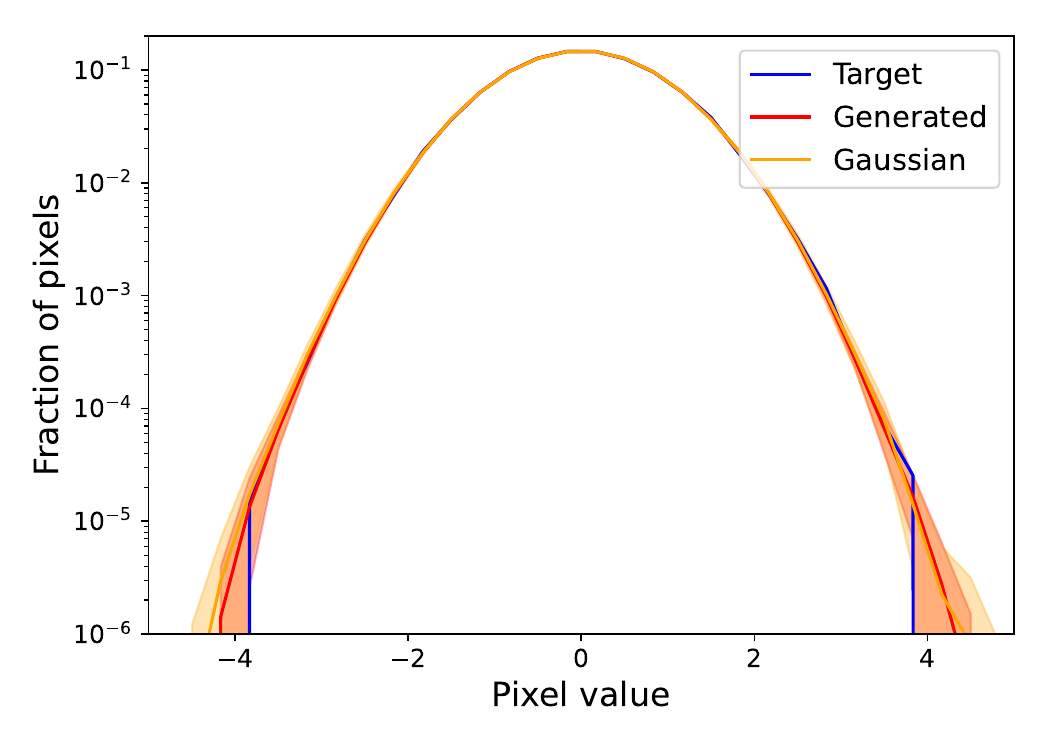}
    \includegraphics[width=0.32\linewidth]{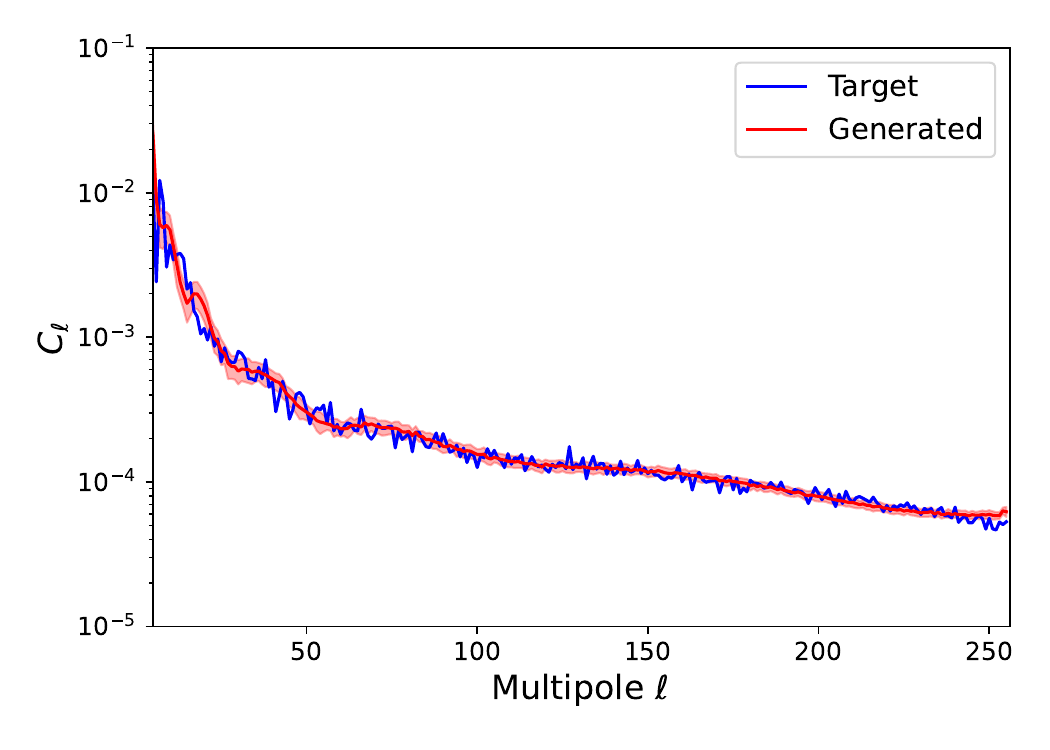}
    \caption{Generative model for a CMB map. The upper part shows (left to right): the target map, a generated field, and a Gaussian realisation. The second row shows the PDF (linear and logarithmic scales) and the angular power spectrum.}
    \label{fig:cmb}
\end{figure*}

It is important to check that the generative model behaves as we expect for a Gaussian field. This is an important validation for a maximum entropy generative model. This is why we tested it on the CMB map. The result is shown in Fig.~\ref{fig:cmb}. The upper parts shows the target map, a generated field and a Gaussian realisation. As expected, the three maps look similar. We also plot the PDF and the power spectrum which match very well. 

\section{Multiple realisations}
\label{appendix_Nreal}

\begin{figure*}
    \centering
    \includegraphics[width=1\linewidth]{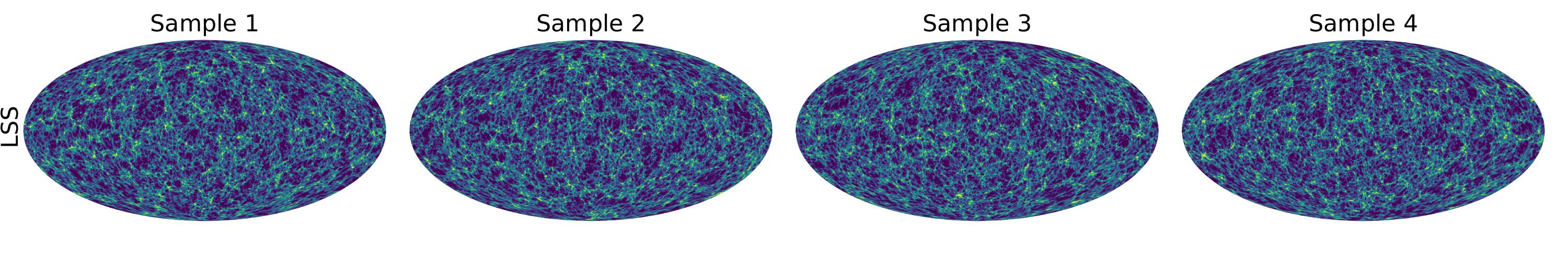}
    \includegraphics[width=1\linewidth]{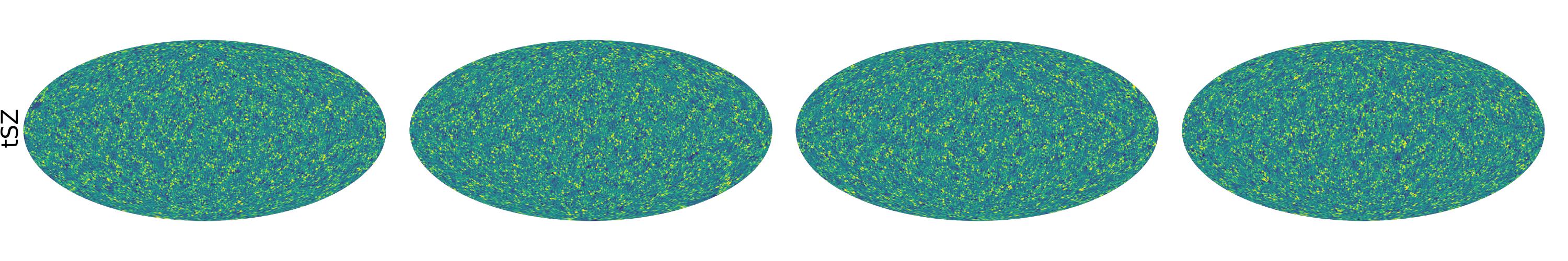}
    \includegraphics[width=1\linewidth]{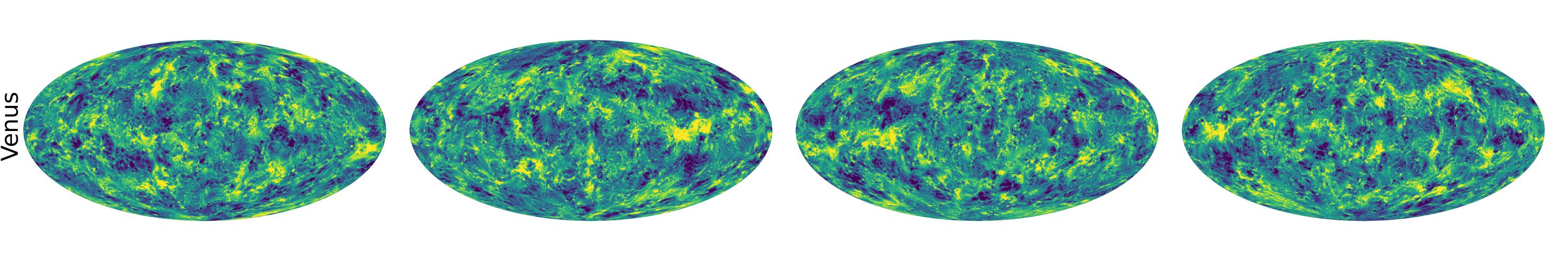}
    \caption{Multiple realisations. Four generative models of the LSS, tSZ, and Venus fields (from top to bottom), obtained by changing the initial Gaussian random noise. In total, we ran 50~realisations for each field. The colour scales are identical within each field.}
    \label{fig:maps_Nreals}
\end{figure*}

Figure~\ref{fig:maps_Nreals} shows multiple realisations obtained from the generative model, changing the initial Gaussian random noise. For each field we show four maps out of the 50 that we computed. This is to illustrate the visual similarity between the realisations.

\section{Minkowski functionals for the three others fields}
\label{appendix_Minko}

\begin{figure*}
    \centering
    \includegraphics[width=0.49\linewidth]{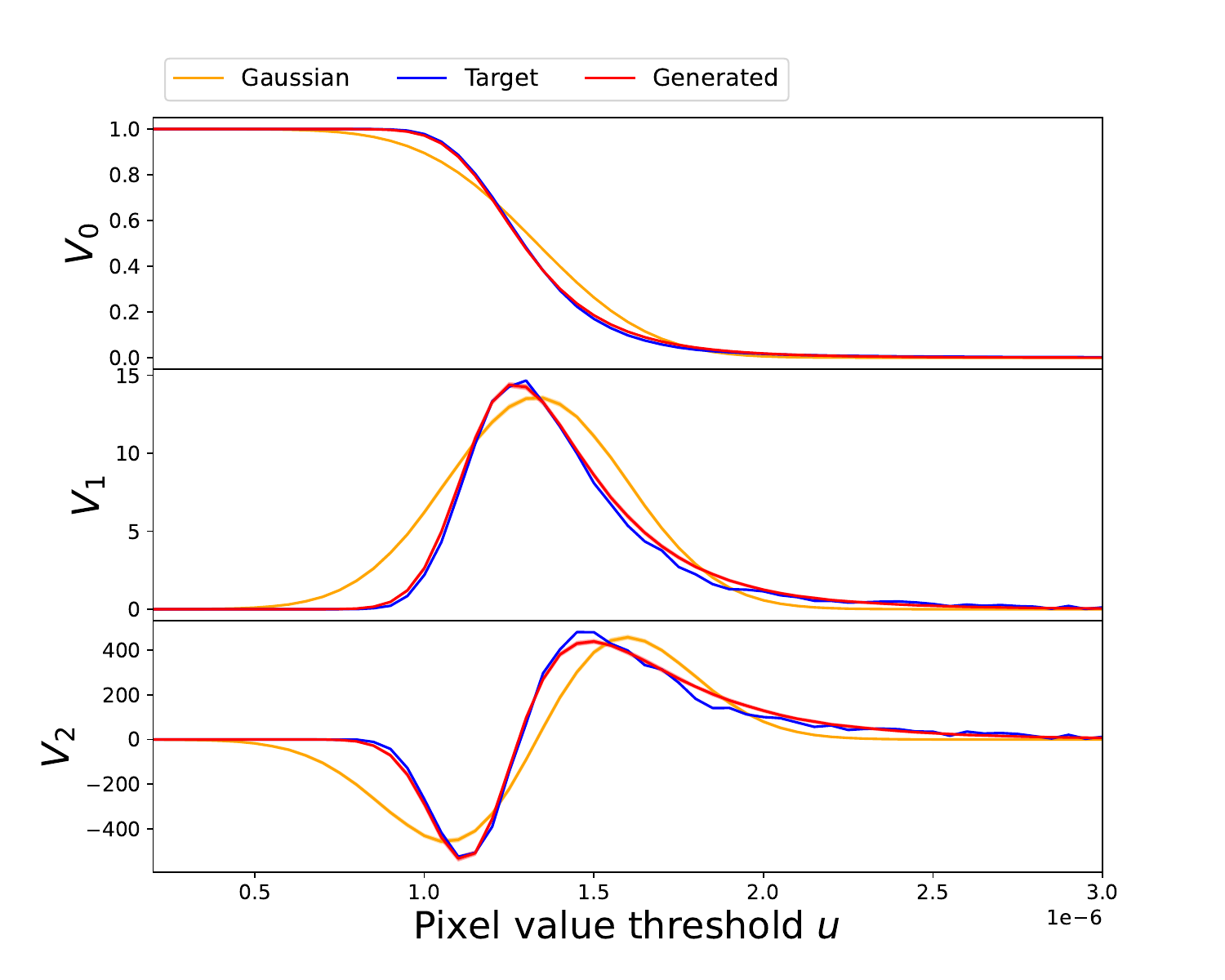}  
    \includegraphics[width=0.49\linewidth]{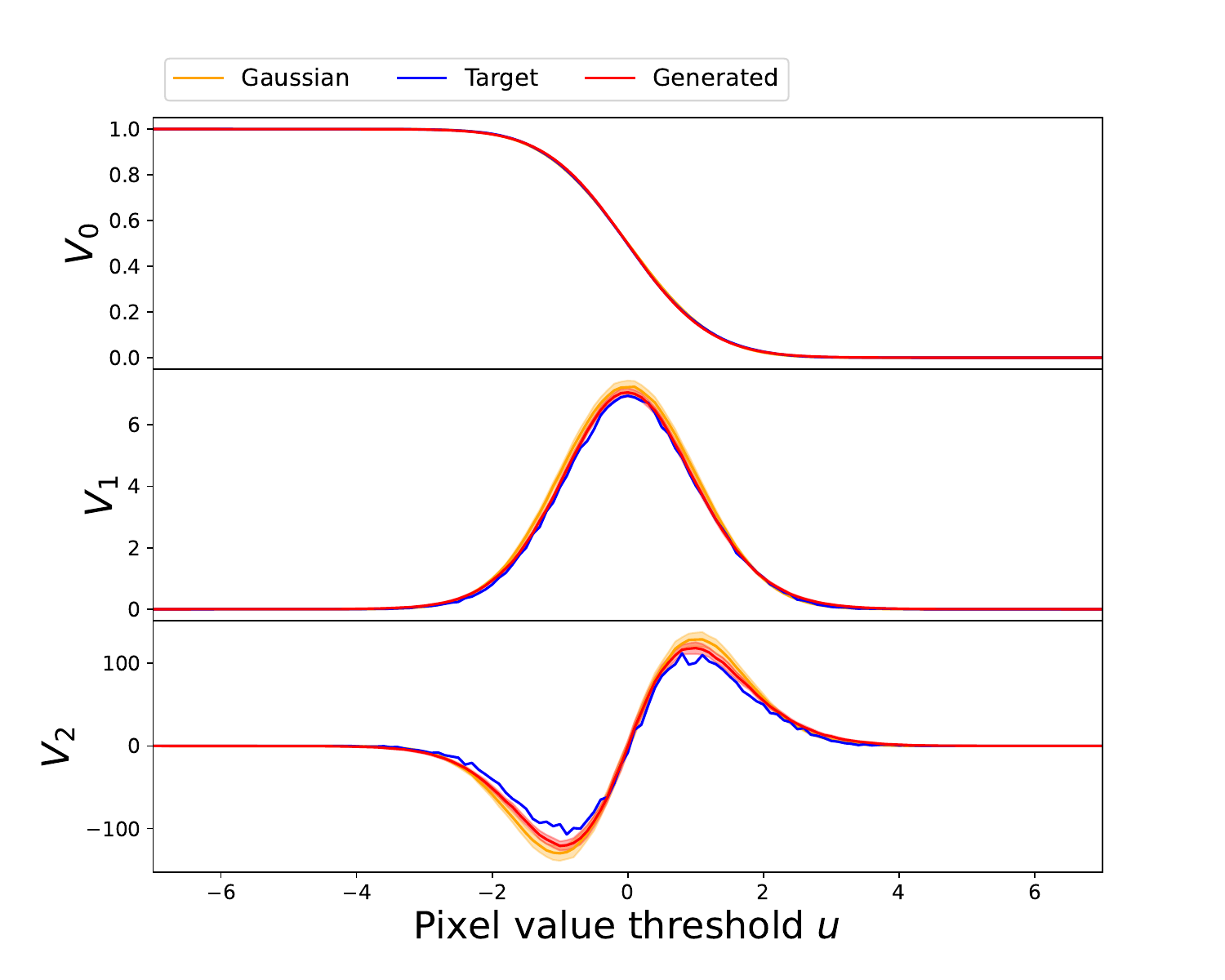}  
    \includegraphics[width=0.49\linewidth]{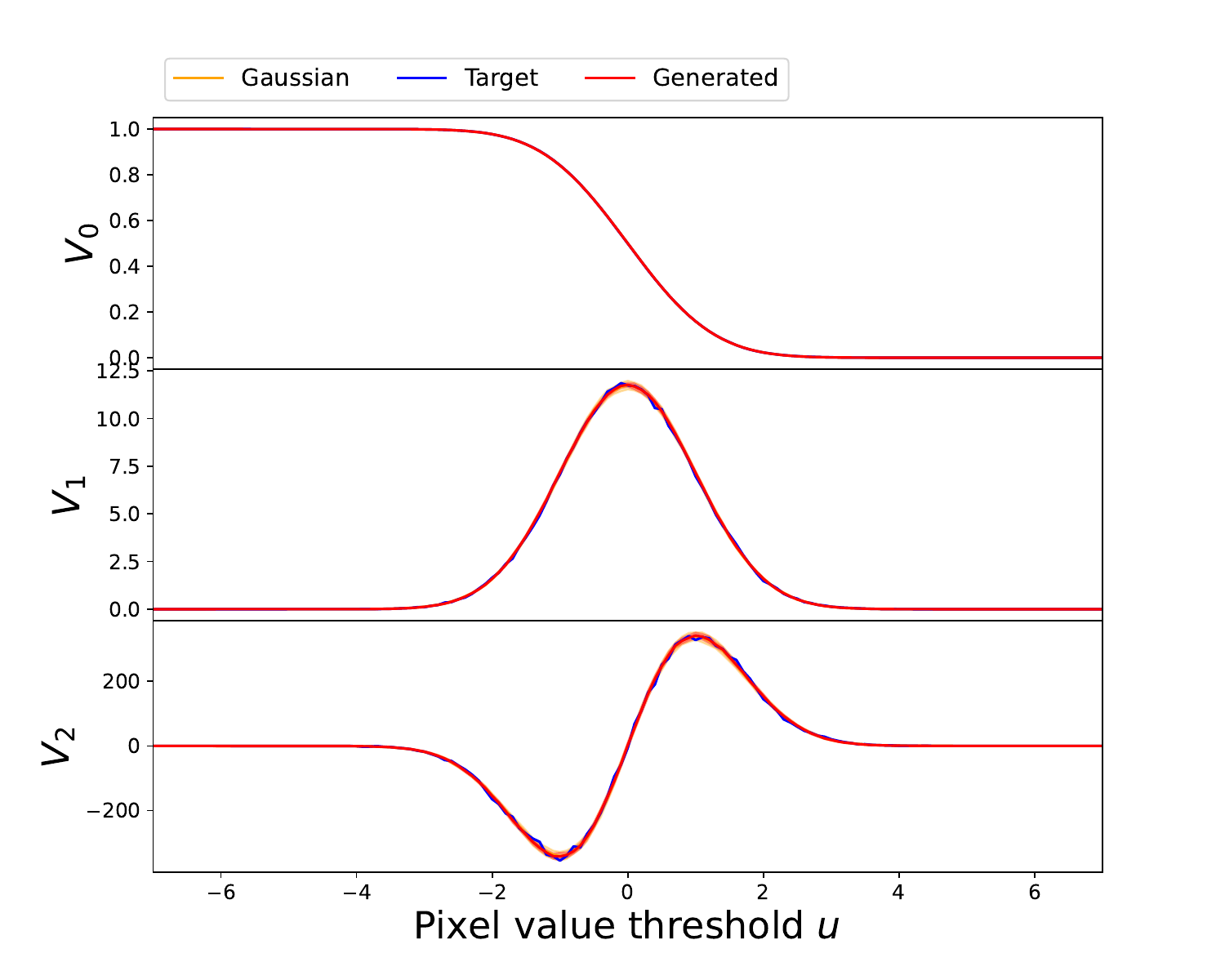}  
    \caption{Additional Minkowski functionals. The three Minkowski functionals $V_0$, $V_1$, and $V_2$ for the tSZ (upper left), Venus (upper right), and CMB (bottom) fields. Blue is the target, red the generated fields, and yellow the Gaussian fields. For the generated fields and the Gaussian realisations, we plot the mean (solid) and the standard deviation (shadow envelope) computed over 50~realisations.}
    \label{fig:Minko_appendix}
\end{figure*}

Figure~\ref{fig:Minko_appendix} shows the Minkowski functionals for the tSZ, Venus, and CMB maps.

\end{document}